\documentclass{article}
\usepackage{amsmath}
\usepackage{lipsum} 
\usepackage[top=2.5cm, bottom=3cm]{geometry}
\usepackage[utf8]{inputenc}
\usepackage{xcolor}
\usepackage{booktabs}
\usepackage{setspace}
\usepackage{graphicx}
\usepackage[hidelinks]{hyperref}
\usepackage[document]{ragged2e}
\usepackage{pythonhighlight}
\usepackage{pdflscape}
\usepackage{longtable}
\usepackage{epstopdf}
\usepackage[flushleft]{threeparttable}
\usepackage{comment}
\usepackage{tikz}
\usepackage{standalone}
\usepackage{array}
\usepackage{pdflscape}
\usepackage{lscape}
\usepackage{adjustbox}
\usepackage{svg}
\usepackage{dutchcal}
\usepackage{changepage}
\usepackage{dutchcal}
\usepackage{caption}
\usepackage{subcaption}
\usepackage{soul}
\usepackage{float}
\usepackage{mdframed}
\usetikzlibrary{mindmap,trees}
\usepackage{stackengine}
\usepackage{afterpage}
\usepackage{natbib}
\usepackage{amssymb}
\usepackage{authblk}
\usepackage{tabularx}
\usepackage{makecell}
\usepackage{enumitem}
\usepackage{multicol}
\DeclareUnicodeCharacter{2064}{}

\title{Can AI Enhance its Creativity to Beat Humans ?\footnote{This work has been funded by the Chair in Management Creativity (Foundation of the University of Strasbourg).}}
\renewcommand{\arraystretch}{1.5}
\date{} 
\author[]{Anne-Gaëlle Maltese\thanks{Corresponding author: ag.maltese@unistra.fr}}
\author[]{Pierre Pelletier}
\author[]{Rémy Guichardaz}
\affil[]{University of Strasbourg, University of Lorraine, CNRS, BETA, 61 Avenue de la Forêt Noire, Strasbourg, 67000, France}

\begin{document}

\maketitle
\justifying

\begin{abstract}
    Creativity is a fundamental pillar of human expression and a driving force behind innovation, yet it now stands at a crossroads. As artificial intelligence advances at an astonishing pace, the question arises: can machines match and potentially surpass human creativity? This study investigates the creative performance of artificial intelligence (AI) compared to humans by analyzing the effects of two distinct prompting strategies (a \textit{Naive} and an \textit{Expert} AI) on AI and across three different tasks (Text, Draw and Alternative Uses tasks). Human external evaluators have scored creative outputs generated by humans and AI, and these subjective creative scores were complemented with objective measures based on quantitative measurements and NLP tools. The results reveal that AI generally outperforms humans in creative tasks, though this advantage is nuanced by the specific nature of each task and the chosen creativity criteria. Ultimately, while AI demonstrates superior performance in certain creative domains, our results suggest that integrating human feedback is crucial for maximizing AI's creative potential.
\end{abstract}

\newpage

\section{Introduction}\label{intro}

Artificial intelligence (AI) is now part of our daily lives in both private and professional spheres. The uses of such technology are manifold, and the fields of application are almost infinite. AI can be defined as ``\textit{a system’s ability to correctly interpret external data, to learn from such data, and to use these learnings to achieve specific goals and tasks through flexible adaptation}'' \citep[p.15]{Kap2019}. Different models and algorithms exist in the realm of AI. Among them are \textit{Transformers}, which are innovative architectures that have radically transformed the field of AI, particularly in Natural Language Processing (NLP) and Image Generation. These models use what are known as ``\textit{attention mechanisms}'' \citep{Vas2017}, an innovative technique that allows the model to focus on different parts of the input sequence when processing or generating each part of the output sequence. Large Language Models (LLMs), such as GPT (Generative Pre-trained Transformer), are a specific class of Transformer models and have become a powerful tool for creators in multiple domains who use algorithms to generate, among other things, works of art, music, writing suggestions, and other similar outputs \citep{Bub2023}. GPTs are now accessible to everyone with widely known user-friendly interfaces such as the one used in this study, \textit{GPT-4} and \textit{DALL-E}.

One of the issues at the heart of our use of AI is its relationship with humans. There are three possible scenarios: humans undertaking a task without AI, humans collaborating with AI, and AI wholly supplanting humans. These scenarios can be relevant for many human activities, including the tasks performed in organisations, as AI has already demonstrated its ability to change them \citep{coc2018, von2018}. Within this context, AI, functioning as a partially autonomous technological entity or agentic technology, has significantly enhanced logistics or decision-making processes. Our study aims to explore another key activity of organisations, namely creativity. Creativity relies on problem-solving and problem-framing, which are essential for organisations as these elements define how organisations operate \citep{Brusoni2005, Miron2018}. As a result, an organisation's creative capacity is a crucial comparative advantage and source of innovation \citep{Chat2018, Woo1993}. Moreover, creative endeavours are highly dependent on the individuals in charge and are highly labour-intensive. Since creativity is highly sensitive to problem-framing and requires substantial investments in human resources, could organisations not consider another path, such as AI? For an organisation to take up the issue of AI as a creative resource, there are two possible options: seeing artificial creativity as a substitute for human creativity or as a complement to it. In other words, AI can be either envisioned as a self-sufficient generator or a complementary creative tool.

The question arises: to what extent does AI compete with human minds when it comes to creative endeavours? While for a long time, the possibility of a machine being creative was dismissed out of hand, new works have emerged to demonstrate that it is a possibility and artificial creativity is now defined as ``\textit{the production of highly novel, yet appropriate, ideas, problem solutions, or other outputs by autonomous machines}'' \citep[p.3]{Ama2019}. As soon as we uphold the premise that creativity is an inherently human endeavour, it opens the door to AI as an alternative to human creativity. With its absence of fatigue, lack of frustration, and faster execution times, AI might be considered a viable substitute. However, the possibility of AI substituting the work of humans might be a cause of concern since the speed of AI advancements might represent a massive and long-lasting displacement of workers. Thus, understanding the performance of AI in creative tasks is critical as organisations increasingly adopt AI technologies, which could change, positively or negatively, the value attributed to human creativity. 

These concerns about humans-AI collaboration or substitution are not new as in the founding work of \cite{Min1961} attesting that ``\textit{With these systems, it will at last become economical to match human beings in real-time with really large machines. This means that we can work toward programming what will be, in effect, thinking aids}'' (p.28). \cite{von2018} put into perspective the stakes deriving from using AI in organisations by considering it as task input, process, and output. It is worth noting that our work focuses mainly on AI as a task output in the sense of the generation of creative solutions. In the end, if the literature on substitution versus complementarity of humans and AI seems to argue for AI as a way to augment human behaviour in organisations, it is still crucial to compare them to understand the specific areas or skills that give rise to this possible complementarity, especially when we know that AI is a fast-growing technology that requires specific attention to track its development and consider the rapidly changing results in the literature. 

Besides the nature of the agent involved in the creative process, comparing outputs requires an assessment of the creative performance of these. Since creativity is a multifaceted concept, it requires a deep understanding of several components to grasp the complexity of it all, and multiple new research avenues are open when it comes to Generative AI.  While the literature on the creative performance of AI is expanding, the variety of tasks tested is still limited and often restricted in the number of dimensions or criteria chosen for the creativity assessment. Moreover, only a few studies have examined the issue of prompting strategies in the specific case of creativity. Indeed, if AI has the capacity to generate ideas, it has no agency. Thus, this generation is always conditional on the request that has been addressed, reinforcing the crucial role played by problem-framing in the creative process. As a result, the research questions at the heart of this work revolve around two major points: the multidimensional nature of creativity and the role of prompting strategies. First, does AI outperform humans in terms of creative performance? On any task? Second, does the prompting strategy adopted to generate creative outputs impact AI creative performance? And what type of prompt engineering strategy proves most effective when aiming for exceptionally high creative performance? 

To do so, an online experiment has been implemented for human evaluators to assess the creative performance of outputs produced by humans and AI. 199 participants rated creative outputs across three tasks (a Text task, a Draw task, and an Alternative Uses task) on multiple criteria. In addition to these evaluations, other measures based on NLP techniques have been used. Our study demonstrates a clearly better performance of AI over humans across all tasks and almost all criteria. However, if AI generally outperforms humans in creative tasks, its success heavily relies on the prompting strategy used. While AI demonstrates strong creative abilities, it faces challenges in more complex tasks without human intervention. Ultimately, the study underscores the importance of human-machine collaboration, as human input, particularly in refining prompts and framing problems, plays a vital role in enhancing the quality and effectiveness of AI-generated creative outputs.

The remainder of the paper is organized as follows: Section~\ref{lit_rev} outlines the state of the art, introducing the relevant literature and concepts evoked in this work. Then, Section~\ref{exp_des} describes the experimental design encompassing the tasks themselves, data generation and collection,  experimental procedure, and hypotheses. Section~\ref{res} presents our study's descriptive statistics, regression analysis, and results. While Section~\ref{disc} and Section~\ref{conc} respectively present the discussion and conclusion of this work.

\section{Literature review}\label{lit_rev}
\subsection{Human and Artificial creativity}\label{lit_rev:crea}

Over centuries, many authors have discussed the origin of creativity. While some began by characterizing creativity as divine inspiration \citep{Gaut2012,Stokes2016}, others followed the romantic view, considering creative individuals as geniuses \citep{Miller1996}. In both approaches, creativity was perceived as something unknown, mysterious, and fundamentally inexplicable. Moving away from this elitist approach to creativity, it has been envisioned as a more complex process and product that can emerge from any individual or group of individuals. Creativity is no longer confined to the scientific or artistic sphere and has been studied from the angle of more ordinary forms of creativity. Creativity can then be simply defined as ``\textit{the ability to come up with ideas or artefacts that are new, surprising and valuable}'' \citep[p.1]{Boden2004}. 

What remains striking in the literature, however, is the inextricable link between creativity and humans, as ``\textit{creativity is part of what makes us human}'' \citep[p.3]{Sawyer2024}. \cite{Ama2016} defines creativity as “\textit{the production of novel and useful ideas by an individual or small group working together}” (p.158). Most creativity-related concepts are articulated around features of the human mind, translating it as a cognitive capacity \citep{Bod1998} that everyone can train. This culminates when creativity is defined as a social process resulting in a bidirectional relationship between individuals and their environment \citep{Ama2016, Csi1996, Fis2005}. However, if creativity is a human-made concept, that should not be a reason for it to remain a human-centric one \citep{Guckel2017}. Indeed, given that creativity is commonly defined as the ability to generate novel and useful ideas \cite{Henn2010}, it seems that any entity capable of generating such novel and useful ideas would be creative. Consequently, new definitions have recently emerged. Artificial creativity is ``\textit{the production of highly novel, yet appropriate, ideas, problem solutions, or other outputs by autonomous machines}'' \citep[p.3]{Ama2019}. Whether using the term artificial or computational creativity, a growing number of authors argue that machines are able to provide human-like creative output using computer systems and algorithms \citep{Veale2019}. 

Since humans and machines are both capable of producing creative outputs, it allows for comparison in performance. Comparison therefore requires a careful examination of how creative performance is assessed. Beyond the duality between humans and machines, one must be careful about the type of task evaluated. Indeed, creativity is a cognitive process, at least if we stick to its human-centric definition, that elicits different thinking processes, mainly categorized as divergent and convergent. While convergent thinking aims to produce a singular logical solution, divergent thinking targets the generation of multiple solutions \citep{Gui1950,Gui1967}. This distinction allows us to differentiate tasks according to their degree of openness, i.e. the degree of autonomy allocated to individuals when carrying out a task. We can then consider three types of tasks based on different degrees of openness: closed (mostly relying on a convergent thinking process and strong constraints), open (mostly relying on a divergent thinking process and few or no constraints), and open with constraints (which is a task closer to real-world circumstances with a mix of divergent and convergent thinking process with some constraints) \citep{Char2019,Att2021}. 

Moreover, creativity is a complex and multifaceted phenomenon that requires a multidimensional approach to its evaluation. Creative performance in a specific task might differ based on the criteria being considered. The literature relies on four main criteria: originality (\textit{how infrequent a particular solution is}), fluency (\textit{how many ideas were generated to solve a specific problem}), flexibility (\textit{how many themes cover the set of ideas generated}), and elaboration (\textit{how detailed an idea is}) \citep{Cas2016, Gui1950, Gui1967}. Besides these traditional criteria, multiple measures of creativity have emerged, also based on the usefulness or appropriateness of ideas that deserve closer attention \citep{Hubert2024} as an idea is a ``good'' idea when it embraces both high originality and high feasibility \citep{Magni2023, Rietzschel2010}. Finally, when assessing the creativity of an idea, one has to question the nature of the evaluator as both objective and subjective measures might be envisioned. While objective assessments are mainly based on quantitative and precise measurements, subjective assessments rely on human evaluators' own personal perception of the output according to the selected criteria. These two assessment methods, although different, are complementary.

In the end, the literature on artificial creativity and its comparison to humans is growing in importance, but the results remain blurred. On the side of task openness, \cite{Charness2024} find a higher performance of humans in an open task, while AI performs better in a closed task, highlighting the complementarity between human and artificial creativity. However, some authors also investigated different criteria of creativity with which the distinction between humans and machines is not straightforward. Even though AI is able to produce human-like outputs, it fails to provide the same level of unexpectedness or novelty of ideas \citep{Stevenson2022}. On the contrary, in \citep{Koivisto2023}, AI outperforms humans in originality and elaboration except for the best of them. Nonetheless, these results need to be nuanced if we look at the technological progress in terms of the AI model at use. While humans were still outperforming GPT-3.5, \cite{Haase2023} observed an almost similar performance between humans and AI when introducing GPT-4. In addition, this difference between AI models is not constant over all selected criteria since GPT-4 seems to outperform GPT-3.5 on fluency but not elaboration \citep{Vinchon2023}. As a consequence, there is no clear-cut on whether AI outperforms humans in creative tasks. For \cite{Haase2023} to conclude that ``\textit{yes} [AI are creative]\textit{, as much or as little as humans}'' (p.11).

\subsection{The specific case of Transformers}\label{lit_rev:trans}

LLM models are a form of transformational models, i.e. statistical models based on the probability of occurrence of the most frequent events. Following this predictive logic, this type of model is trained on a large ensemble of data and, based on the various inputs and the most frequently occurring events previously generated, provides a coherent and appropriate response. These types of model can be referred to as GPTs (Generative Pretrained Transformer), among which the well-known application from OpenAI is GPT-4. Released in March 2023, GPT-4 aims to generate human-like texts depending on the context, particularly in response to a prompt. Then, DALL-E 2 is an AI image generation tool developed by OpenAI in 2022 \citep{ramesh2022hierarchical}, which generates an image based on a text description. The system underlying DALL-E 2 is based on two technologies, namely CLIP and diffusion. CLIP, which stands for Contrastive Language-Image Pre-training, plays a crucial role in DALL-E 2, serving as the main bridge between text and images. ⁤⁤It consists of two neural networks: a text encoder and an image encoder, both trained on a large dataset of image-text pairs. ⁤⁤These encoders map their inputs into a shared vector space, effectively creating a ``concept space'' which allows for the translation of semantic information between text and images. ⁤In DALL-E 2, the diffusion model complements CLIP by generating images based on the embeddings produced by CLIP. CLIP first processes the input text to create a text embedding, which is then used to generate a corresponding image. This embedding vector captures the semantic meaning of the text in a form that can be used for image generation. The diffusion model then takes the embedding and generates an image by iteratively refining noise into a coherent visual representation.

Based on attention mechanisms, GPTs aim to provide answers for a given context and serve as problem-solving tools. Problem-solving requires looking for solutions in a solution space using reasoning that begins with the formulation of the problem to apply it to the information acquired by the solver where there is no pre-determined operator or strategy \citep{Ollinger2010, Simon1973}. Additionally, problem-solving is closely related to problem-framing. When evaluating AI's problem-solving capabilities, it is crucial to consider how the problem is formulated and how this formulation impacts the resolution. In the case of Generative AI, this means looking at the exact request formulated by the user and addressed to the model. This request is referred to as a prompt, where ``\textit{prompting means prepending instructions to the input and pre-training the language model so that the downstream tasks can be promoted}'' \citep[p.3]{Ge2023}. Since AI lacks agency, making its creative potential reliant on human assistance to generate responses \citep{Hubert2024}, the use of prompts becomes the main driver for problem-solving. 

This prompt-dependency justifies the need to reflect on what prompting strategy should be selected by users. Indeed, any user of LLMs has experienced that the obtained output is highly dependent on what they have asked for and that some changes in the phrasing or framing of a question or instruction might totally change the result. As a consequence, new skills have emerged in organisations and for users of AI that need now to excel in prompt engineering, i.e., the process of creating, refining, and optimizing prompts. In fact, you might prefer to ask a question directly to the AI interface without any context or give a complete and comprehensive context to the interface to ensure that AI enhances its creative capacities. In other words, one might choose different prompting strategies using LLMs when trying to maximize AI performance depending on the intended objective. As for any relevant phenomenon, literature has emerged on the concept of prompting strategies, highlighting the importance of adapting the prompt when it comes to specialized tasks \citep{Baidoo2023}. These strategies encompass prompts based on personality \citep{Chen2023, Xu2022}, on a context for a specific field \citep{Ge2023}, or Chain-of-Thought reasoning \citep{Wei2022}. However, no clear consensus prevails on the effect of each strategy as results might differ according to the elicited strategy as well as the chosen task. 

While we recognize that AI is increasingly capable of solving ever more complex problems \citep{von2018}, we must also acknowledge a number of shortcomings. In fact, the use of AI does not guarantee a zero error rate as soon as we admit that models can suffer from hallucinations. These hallucinations refer to unreliable and nonsensical text, image, audio, or video outputs generated by LLMs \citep{Rawte2023}. In these cases, prompts again play a central role as they contribute to \textit{dehallucinating} them as the inaccuracy of generated outputs decreases. However, hallucinations might be valuable when used for creative purposes. Indeed, as sources of unexpected answers, hallucinations might fuel creativity \citep{Rawte2023}. While this change of perspective may seem curious, it nonetheless calls into question the link between AI and creativity when we consider the points made earlier about prompt dependency and the resulting prompting strategies. Thus, besides the earlier mentioned concerns on the criteria chosen to assess creative performance, it is also essential to question what determines an ideator's creative performance when it comes to problem-framing. This study aims to tackle this question by comparing two different prompting strategies characterized as \textit{Naive} and \textit{Expert} prompting while comparing different types of tasks and the resulting creativity criteria.

\section{The experiment}\label{exp_des}

The description of our experimental protocol is divided into two main parts. The first part is dedicated to the collection and generation of creative outputs by humans and AI. The second part implements an online experiment dedicated to the evaluation of these outputs by human evaluators.  

\subsection{Data collection and generation}\label{exp_des:data}

The creative outputs used in this study can be broken down into two components: experimental data produced by humans and AI-generated data. In the case of human data, these were collected through three different creativity tasks conducted during two experimental protocols \citep{Malt2023, Guichardaz2024}. Table~\ref{tab:crea_crit} summarises the tasks and their characteristics.

\vspace{0.3cm}
\begin{table}[H]
    \centering
    \small
    \caption{Tasks' Characteristics}
    \begin{tabularx}{\textwidth}{ p{3cm} >{\raggedright\arraybackslash}p{5cm} p{3cm} p{3cm} }
    \toprule
    \midrule
    Tasks & Aim & Nature of the task & Main thinking process involved \\
    \midrule
        Text task & \makecell[l]{Writing a text based on a \\ list of predefined and \\ compulsory words} & \makecell[l]{Close-ended} & Convergent \\
        \addlinespace
        Alternative Uses task & \makecell[l]{Finding unusual uses \\ for everyday objects} & \makecell[l]{Open-ended \\ with constraints} & Divergent \\
        \addlinespace
        Draw task & Draw an alien animal & Open-ended & Divergent \\
    \midrule
    \bottomrule
    \end{tabularx}
    \label{tab:crea_crit}
\end{table}
\vspace{0.3cm}

This sequence of tasks aims to provide a more comprehensive understanding of creative contexts than what is found in the existing literature, particularly by capturing a broader spectrum of both convergent and divergent thinking processes, as well as varying degrees of task openness encompassing not only text tasks but also a drawing one. To start with, the Text task inspired by \cite{Char2019} is a closed task that requires the subject to combine a set of elements acting as constraints, providing a clearer goal to reach. As a result, such a task elicits more of a convergent thinking process. This Text task allows us to observe the capacity of humans and AI to produce a creative text while thematic constraints have been introduced. 

Next, the Draw task aims to evaluate subjects' creativity through an open task. Introduced by \cite{War1994}, this task was specifically chosen to represent and elicit a divergent thinking process. While selecting a common idea requires less cognitive effort, the additional effort of exploring unusual and original ideas enhances creative performance. \cite{War1994, War2004} refers to this as the ``path of least resistance''. In this task, when individuals are asked to imagine and create an extraterrestrial animal, they often incorporate many terrestrial characteristics (such as bilateral symmetry or sensory organs) into their creations \citep{War1994, War1997}. This demonstrates a fixation on familiar animal concepts. However, more creative individuals distinguish themselves by moving beyond these familiar elements, though ``\textit{the ability to generate a creative idea begins with known concepts}'' \citep[p.45]{Birdsell2019}.

Finally, the Alternative Uses Task requires subjects to find unusual applications for ordinary items \citep{Tor1966, Gui1967}. This task falls under the category of an open-with-constraints task. Widely employed for evaluating creative capacities, the Alternative Uses task is credited as one of the most prevalent methods. The aim of such a task is to assess individuals' capacity to generate a multitude of novel ideas, relying on more divergent thinking processes while requiring a certain degree of usefulness or appropriateness of ideas, which requires an ability to converge towards feasible ideas. 

The same tasks were then used to produce the AI creative outputs. These were generated using GPT-4 (Text and AUTs), coupled with DALL-E 2 (Draw task). More specifically, AI was prompted to perform the same tasks as our human subjects, with the only difference being the use of two different prompting strategies. Firstly, an AI whose prompts are as neutral as possible, with no other indications than the creative task itself (\textit{naive prompting}). Secondly, an AI that we prompted so that it might be more effective in the specific task (\textit{expert prompting}). The purpose of these two strategies is to capture the two extreme cases of human use of AI. The \textit{Naive} AI corresponds to a minimalist use of AI, akin to what an uninformed individual in the field might use. Conversely, \textit{Expert} AI corresponds to the situation where a user employs a more sophisticated approach to prompt, thereby pushing its capabilities further. 

Throughout the rest of the paper, we will refer to the outputs as \textit{Human}, \textit{Naive AI}, and \textit{Expert AI} to differentiate them. Appendix~\ref{app:exp_instr} presents the instructions for each task and the related pre-prompt in GPT-4 for both strategies.

\subsection{Creativity assessment}\label{exp_des:crea_ass}

Once the creative outputs have been collected or generated, the next step is to assess the creativity of these outputs. Outputs are evaluated according to the specific task and the corresponding creativity criteria as presented in the literature. 

To start with, evaluators were randomly assigned to pools to evaluate one specific task among the three described above. This choice of separating evaluators by task has been made in order to avoid contamination from one type of evaluation to the other. Additionally, due to the difference in cognitive load to evaluate each task, evaluators were presented with either 15 outputs from the Text task, 15 from the Draw task, or 9 from the Alternative Uses task (where one output represents a list of ideas for one specific word). AUT, in particular, demanded evaluators to assess each idea individually, rendering the overall evaluation process more laborious compared to the other tasks. In the end, each output has been evaluated by at least 2 randomly matched evaluators from the assigned pool (1.32\% by 2, 27.34\% by 3, 71.36\% by 4 evaluators).

Before proceeding with their scoring, the evaluators are provided with the instructions given to the creators of the outputs (whether human or AI) as presented in Appendix~\ref{app:exp_instr}. Then, each evaluator assigns a score ranging from 0 to 5 to each creativity criterion. Evaluators were not informed at any point that some of the outputs they were assessing might be generated by AI. They were simply asked to evaluate each output individually based on the creativity criteria outlined in Table~\ref{tab:crea_crit}. 

\begin{table}[H]
    \centering
    \caption{Creativity Criteria Assessed by Evaluators}
    \scalebox{0.85}{
    \begin{tabular}{llccc}
    \toprule
    \midrule
        Criteria & Definition & Text & Alternative Uses & Draw \\
    \midrule
        Validity & \textit{Adherence to instructions} & X & X & X\\
        Form & \textit{Style and writing or drawing quality of outputs} & X & & X \\
        Elaboration & \textit{Degree of details} & X &  &X \\
        Originality & \textit{Unusualness of ideas}& X & X & X\\
        Feasibility & \textit{Viability of ideas} &  & X & \\
    \midrule
    \bottomrule
    \end{tabular}}
    \label{tab:crea_crit}
\end{table}

To conclude the experimental sessions, evaluators answered a final questionnaire encompassing sociodemographic information and questions related to AI detection, usage, and attitudes. The questionnaire is available in Appendix~\ref{app:quest}.

\vspace{0.5cm}

\subsection{Experimental procedure}\label{exp_des:proc}

The experimental protocol was conducted online. In total, 199 evaluators\footnote{Only two evaluators were removed from the original pool of 201 evaluators because they dropped out before completing the experiment.} were recruited via the ORSEE platform from the LEES, the experimental economics lab at the University of Strasbourg, which primarily consists of student email addresses. As evaluators only evaluated specific tasks, 65 of them evaluated the Draw task, 70 the Text task, and 64 the AUT. Evaluators received a flat payment of €15 for their participation, and the average total response time for completing the experiment was 27 minutes. Table~\ref{tab:stat_desc} presents some descriptive statistics about the evaluators' sociodemographic information, and Table~\ref{tab:comp_stat_desc} shows the comparison of populations of evaluators across tasks. 

\vspace{0.5cm}
\begin{table}[!htbp] \centering 
\renewcommand{\arraystretch}{1}
\setlength{\tabcolsep}{0.2pt} 
  \caption{Evaluators' Socio-Demographic Characteristics per Task}
  \label{tab:stat_desc} 
\scalebox{1}{
\begin{threeparttable}
\begin{tabular}{@{\extracolsep{5pt}}lccc} 
\\[-1.8ex]\hline 
\hline \\[-1.8ex] 
Variable & Text & Alternative Uses & Draw\\
\midrule
Mean Age & 22.814 (3.036) & 22.094 (2.646) & 22.031 (2.952)\\
Mean Gender & 0.657 (0.475) & 0.672 (0.47) & 0.831 (0.375)\\
Mean Diploma Licence & 0.414 (0.493) & 0.484 (0.5) & 0.415 (0.493)\\
Mean Diploma Master & 0.443 (0.497) & 0.469 (0.499) & 0.508 (0.5)\\
Mean Diploma Doctorat & 0.057 (0.232) & 0 (0) & 0.031 (0.173)\\
Mean Droit & 0.057 (0.232) & 0.047 (0.212) & 0.015 (0.123)\\
Mean Economie Gestion & 0.414 (0.493) & 0.297 (0.457) & 0.4 (0.49)\\
Mean Lettres Langues & 0.057 (0.232) & 0.047 (0.212) & 0.046 (0.21)\\
Mean Sciences exactes & 0.186 (0.389) & 0.203 (0.403) & 0.154 (0.361)\\
Mean Psycho Socio & 0.071 (0.258) & 0.062 (0.242) & 0.092 (0.29)\\
Mean Sciences politiques & 0.029 (0.167) & 0.094 (0.292) & 0.092 (0.29)\\
Mean Native speaker & 0.8 (0.4) & 0.906 (0.292) & 0.923 (0.267)\\
Mean French skills & 7.671 (1.481) & 8.25 (1.415) & 8.462 (1.039)\\
\hline
\# Evaluators & 70 & 64 & 65\\
\hline 
\hline \\[-1.8ex] 
\end{tabular} 
 \end{threeparttable}
 }
\end{table} 

\vspace{0.5cm}
\begin{table}[!htbp] 
\centering 
\renewcommand{\arraystretch}{1}
\setlength{\tabcolsep}{0.2pt} 
  \caption{Comparison of Evaluators' Socio-Demographic Characteristics Across Tasks} 
  \label{tab:comp_stat_desc}
\scalebox{1}{
\begin{threeparttable}
\begin{tabular}{@{\extracolsep{5pt}}lccc} 
\\[-1.8ex]\hline 
\hline \\[-1.8ex] 
Metric & Draw vs Alternative Uses & Text vs Alternative Uses & Text vs Draw\\
\midrule
\addlinespace
Age & -0.063 & \textcolor{blue}{0.721}$^{***}$ & \textcolor{blue}{0.784}$^{***}$\\
Gender & \textcolor{blue}{0.159}$^{***}$ & -0.015 & \textcolor{red}{-0.174}$^{***}$\\
Native speaker & 0.017 & \textcolor{red}{-0.106}$^{***}$ & \textcolor{red}{-0.123}$^{***}$\\
French skills & \textcolor{blue}{0.212}$^{***}$ & \textcolor{red}{-0.579}$^{***}$ & \textcolor{red}{-0.79}$^{***}$\\
\addlinespace
\textbf{Diploma} & & & \\
Bachelor & \textcolor{red}{-0.069}$^{**}$ & \textcolor{red}{-0.07}$^{**}$ & -0.001\\
Master & 0.039 & -0.026 & \textcolor{red}{-0.065}$^{**}$\\
PhD & \textcolor{blue}{0.031}$^{***}$ & \textcolor{blue}{0.057}$^{***}$ & \textcolor{blue}{0.026}$^{***}$\\
\addlinespace
\textbf{Domain} & & & \\
Law & \textcolor{red}{-0.031}$^{***}$ & 0.01 & \textcolor{blue}{0.042}$^{***}$\\
Economics and management & \textcolor{blue}{0.103}$^{***}$ & \textcolor{blue}{0.117}$^{***}$ & 0.014\\
Humanities & -0.001 & 0.01 & 0.011\\
Exact sciences  & \textcolor{red}{-0.049}$^{**}$ & -0.017 & 0.032\\
Psychology and sociology & \textcolor{blue}{0.03}$^{*}$ & 0.009 & -0.021\\
Political sciences & -0.001 & \textcolor{red}{-0.065}$^{***}$ & 
\textcolor{red}{-0.064}$^{***}$\\
\addlinespace
\hline 
\hline \\[-1.8ex] 
\end{tabular} 
\begin{tablenotes}
\footnotesize
\item \justifying {\it Notes:} This table presents the results of the ANOVA and Tukey's HSD test for the specified variable and group. The coefficients represent the Mean differences between groups. ***, **, and * indicate significance at the 1\%, 5\%, and 10\% levels, respectively.
\end{tablenotes}
 \end{threeparttable}
 }
\end{table} 

\section{Hypotheses}\label{hyp}

Our study aims to compare the creative performance of AI and humans and how this performance is influenced by the nature of the creative task at hand and the chosen prompting strategy for creative outputs generated by AI. We formulate three hypotheses based on the literature in Section \ref{lit_rev} and stylized facts. 

Unlike humans, AI operates without mental or physical fatigue and can generate and manipulate a vast array of elements to achieve its objectives. We might consider, for instance, the contrast in time investment between AI and humans to compose or revise a text or create a drawing. With elements such as grammar, syntax, and semantics, AI demonstrates a significant ability to produce relevant and coherent responses that align with the formal requirements dictated by linguistic conventions (or visual conventions in the case of DALL-E). Nonetheless, it is necessary to consider the quality of human and artificial production in more detail, depending on the task in question. 

As noted earlier, transformer-based models excel in processing vast amounts of data and identifying patterns within this data, enabling them to generate precise, contextually appropriate responses to specific prompts. Indeed, transformer-based models allow to ``\textit{access and generate larger amounts of knowledge, which in turn results in more possible connections of problems and solutions}'' \citep[p.142]{Bouschery2023}. This proficiency suggests that AI systems may outperform humans in close-ended tasks, which benefit from the model's ability to identify the most accurate or optimal solution based on its learned patterns.

However, creativity is not just a matter of (re)combination but also novelty. So, although the knowledge set might be larger for an AI, the ability to combine distant elements is not unique to the latter. Indeed, in its configurations, a Generative AI model aims to generate relevant outputs, playing on the prediction of probable and closely connected consecutive elements. Some authors criticized these AI productions as they ``\textit{merely emulate cognitive processes and cannot substitute the great flexibility, adaptability, and generativity we associate with human intelligence}'' \citep[p.408]{von2018}. Conversely, human cognitive processes are more inclined to explore novel ideas in drawing on personal experiences, emotions, and cross-disciplinary knowledge, aspects where AI currently shows limitations. Indeed, even though AI and, more specifically, transformer-based models provide a broader set of knowledge and possible combinations \citep{Bouschery2023}, the originality of ideas always seems to be in favour of human creativity \citep{Koivisto2023}. Even though GPTs' hallucinations do exist, ``\textit{given that LLMs are designed to generate approximately the statistically most plausible sequence of text based on their training data, perhaps they generate less-novel ideas}'' \citep[p.7]{Gir2023}, which might harm their ability to perform in more open-ended tasks oriented towards a more divergent thinking process. Therefore, we propose the following hypothesis: 

\vspace{10pt}
\textit{\textbf{Hypothesis 1}: AI outperforms humans in close-ended tasks, whereas humans outperform AI in open-ended tasks.}
\vspace{10pt}

Then, if we consider the comparative performance of humans and AI, we must also consider the possible effect of the prompting strategy on the performance of AI itself. The literature has recently emphasized the crucial role of prompting strategies in augmenting the performance of large language models. Prompt serves as a navigational tool, enabling users to steer the model towards desired outputs by providing structured cues and directives. Moreover, orienting prompts offer users a mechanism to refine model responses through the strategic incorporation of specific instructions, constraints, context, or examples. This strategic tailoring not only ensures the alignment of model outputs with intended objectives but also enhances model performance across diverse tasks and domains. In the context of this paper, prompting strategies refer to the \textit{Expert} use of AI compared to the \textit{Naive} use of AI to differentiate between a Generative AI model designed to perform well in creativity tasks and a more neutral model. Therefore, we formulate the following hypothesis:

\vspace{10pt}
\textit{\textbf{Hypothesis 2}: Expert AI outperforms Naive AI across all tasks and creativity criteria.} 
\vspace{10pt}

\section{Methodology}\label{hyp}

\subsection{Output Generation}\label{subsec:output_gen}

As described in section \ref{exp_des:data}, our agents are of three types: Human, Naive AI and Expert AI. In this section, we describe the instructions given to GPT-4 to generate the two types of AI agents (Details of the instructions given have been translated in appendix \ref{app:exp_instr}). GPT-4 can be queried via the API by submitting both a user prompt and a system prompt. The system prompt is used to give initial instructions or directives that help to shape the model's responses. For the naive AI, we simply submitted a user prompt to GPT-4 with the instructions given to each human (most similar to the use of GPT-4 on the OpenAI platform). To create the expert AI, we proceeded in two steps: the first was to provide GPT-4 with the instruction, asking the AI to generate a system prompt that would amplify its creativeness. The system prompt generated by GPT-4 will then complement the user prompt, which recalls the exercise instruction, enabling GPT-4 to respond to the instruction while being already set on a creative path. Of note, the creative ouput generation was calibrated at a temperature of 0.9 to maximize the coherence of answers while minimizing their possible redundancy.

\subsection{Metrics}

To compare our different agents, we used three types of measurement. The first type is human-based and was described in section \ref{exp_des:crea_ass}. The other two types are built using GPT-4 or the vector representation of different outputs and are detailed below.\\

\textit{\textbf{Theme-based metrics}}: We constructed measures of variety, balance, diversity, minimum theme frequency and uncommonness of theme combination scores across all tasks when applicable. For the AUT, we gave GPT-4 the set of individuals' responses for each word and requested it to create a list of 25 themes for each of the words. We then gave GPT-4 each individual idea again, along with the 25 themes, and asked it to assign the most relevant category to each idea. For the Text task, we proceeded differently, as we couldn't give GPT-4 all the texts simultaneously. We, therefore, first gave each text to GPT-4 to assign several themes, then retrieved all these themes and asked GPT-4 to group them together to give us a total of 25 possible themes. Finally, we resubmitted each text in GPT-4 and asked it to assign the most pertinent themes to each text. In this way, we were able to construct three measures based on these themes, namely variety (the number of themes as a measure of \textit{flexibility}), balance (the proportion with which themes are used in an individual's responses as a measure of fixation effect,\footnote{Here, we took the standard deviation of the proportion; a value of 0 means that all themes are uniformly distributed.} the diversity is expressed by the Shannon index, which considers both the richness (how many distinct themes) and evenness (how equally the ideas are distributed across these themes). We also computed the proportion with which themes are used in all responses and kept the minimum for each observation (Theme Frequency (neg.)\footnote{We present this proportion as negative for readability; a higher value means that the usage of the given theme across all observations is low.}). Finally, we calculated an uncommonness score \textit{à la} \cite{lee2015} to provide a metric on the combination of themes.\footnote{This indicator captures the ratio of the observed number of co-occurrences to the expected number and shows how the pairing of two concepts is unusual or common.} \\

\textit{\textbf{Embedding-based metrics}}: Other quantitative measures are based on the embedding of texts, ideas or images. We represented our textual data in a semantic space (separating both tasks) using embedding techniques based on transformers. Rather than employing embeddings through OpenAI's API, we used state-of-the-art French text representation, i.e. CamemBERT,\footnote{CamemBERT is based on the Transformer architecture, which enables it to process texts bidirectionally and capture context efficiently. It is a BERT model trained on the French part of OSCAR (Open Super-large Crawled Aggregated Corpus). More specifically, we use specialized French sentence embedding models such as Sentence-CamemBERT-Large, which can represent the semantics and meaning of French sentences in the form of mathematical vectors \citep{reimers2019sentence, martin2020camembert}. Using this pre-trained model, we can project our sentences into a 768-dimensional vector space and compare them by calculating semantic distances using cosine similarity (Cosine Distance).} as, in general, it is preferable to use language-specific models rather than multilingual models to represent texts. For both Text and AUT, we have broken down texts into sentences or specific ideas, respectively giving each sentence in a text and each idea in a list its own representation in the task-related semantic space. This strategy allows us to understand how an individual agent (Human or AI) uses concepts that are distant from each other when responding to an instruction. We opted for this solution as a representation of all the ideas or texts of a specific agent would not reflect the extent to which the agent exploits the knowledge space. In a similar way, we represented images in a vector space using the vision transformer model called CLIP. In this way, we can also project the drawings of different agents into a space that captures similarities at the visual level,\footnote{As explained in Section \ref{lit_rev:trans} this model mainly allows text and images to be projected into a similar space, but we focus here on the visual part.}. Although this doesn't allow us to construct distance measures at the level of drawings, it does enable us to capture distances between drawings and thus understand the heterogeneity of responses given by different agents. We calculated a centroid for each agent type and compared the distances of the different outputs to this centroid (Cosine Distance to centroid). Due to their high dimensionality, these vectorial representations cannot be visualized directly. Therefore, in order to understand whether our texts exist in the semantic space in a somewhat different way, we have represented them graphically by reducing the dimensionality of this semantic space. A typical technique to reduce the dimensionality of a semantic space is to use t-Distributed Stochastic Neighbor Embedding (T-SNE)\footnote{T-SNE  is a technique used to visualize and understand high-dimensional data sets and is very effective at preserving the local structure of the data, meaning that points that are close in high-dimensional space remain close in low-dimensional space. Unlike linear methods (such as PCA), t-SNE can capture non-linear relationships within the data.} \citep{van2008visualizing}. \\

\textit{\textbf{Controls}}: Finally, we created control variables for each of the tasks, namely the number of words for the text task, the average number of words per agent idea for the AUT, and the proportion of black pixels present in each image.

\subsection{Analysis}

Lastly, our analysis is based primarily on comparing the means of the various metrics between the different agents. First, we performed an ANOVA with a parametric Tukey's HSD test, which we complemented with a non-parametric Pairwise Wilcoxon test to compare all the metrics (human-scored, theme-based, embedding-based and our control variables). The measures given by humans are Likert scales. Therefore, we then performed ordered polynomial logit regressions to understand how agent type plays on performance in each of the dimensions, taking into account our control variables. Finally, we also investigated how evaluators' socio-demographic characteristics influenced their responses and used an ordered polynomial logit as well.

\section{Results}\label{res}

The results of our study are divided between the three creative tasks submitted for evaluation, incorporating both objective and subjective measures, \footnote{In addition, a principal component analysis, the results of which are available in Appendix~\ref{app:pca}, was carried out, demonstrating that our objective and subjective measures do indeed represent distinct elements of creative performance.}. Table~\ref{tbl:res_summary} summarises the results obtained for our objective and subjective measures. \\

\begin{table}[h!]
\caption{Summary of Regression Results}\label{tbl:res_summary}
\hspace*{\fill} 
\resizebox{0.85\textwidth}{!}{%
\begin{tabular}{lccc}
  \toprule
  \midrule
  & Text task & Alternative Uses task & Draw task \\ 
  \midrule
  \textbf{\textit{Objective measures}} &   &   & \\ 
  Variety & \(N > E = H\)  & \(N > E > H\)  & \(H > N* \)\\ 
  Balance & \O &  \textit{no significance} & \O\\ 
  Diversity & \O &  \(N > E > H\) & \O\\ 
  Theme Freq. & \textit{no significance} &  \(E = N > H\) & \(H > N*\)\\ 
  Uncommonness & \(N = H > E\) &  \( E = N > H\) & \(N > H*\)\\ 
  Cosine distance & \(H > E > N\) & \textit{no significance}  & \O\\ 
  Distance to centroid & \(H > E > N\) & \(H > E = N\)  & \(H > N > E\)\\
  \addlinespace
  \midrule
  \textbf{\textit{Subjective measures}} &   &   & \\ 
  Validity & \(N = H > E\) &  \(E = N > H\) & \(E > N = H\)\\ 
  Form & \(N > E = H\)  &  \O & \(E > N > H\) \\
  Elaboration & \(N > E = H\)  & \O & \(E > N > H\) \\ 
  Originality &  \(N = H > E\) &  \(E = N > H\) & \(E > N = H\) \\ 
  Feasibility & \O  & \(E > N > H\)  & \O\\
  \midrule
  \bottomrule
\end{tabular}%
}
\hspace*{\fill} 
\begin{tablenotes}
 \footnotesize
 \centering
 \item * signals that there were no significant differences between Expert AI and the two other agents,\\ 
 Naive AI and humans. \O~ indicates that the criterion is not computed or evaluated for this \\
 specific task.
\end{tablenotes}
\end{table}

\vspace{1cm}

\subsection{Text task}\label{res:txt}

As a reminder, the Text task requires agents to write creative text based on a list of mandatory words. The objective measures of creative performance, presented in Table~\ref{tbl:txt_tukey}, show mixed results on whether AI outperforms humans or not in terms of performance. First, Naive AI outperforms Expert AI and humans in terms of \textit{Variety} (the number of themes), while there is no significant difference between Expert AI and humans. We do not find any significant effect on \textit{Theme frequency}. But, for the \textit{Uncommonness} of theme combinations, Naive AI performs similarly to humans, and both outperform Expert AI. Finally, in terms of \textit{Cosine distance}, humans outperform Expert and Naive AI, and Expert AI outperforms Naive AI. Taken together, these objective measures tell us something about the creative performance of our three agents, and in particular, that none of them is clearly outperforming. We observe that Naive AI is able to produce texts with a higher number of different themes but also to provide combinations of themes that are more unusual. While humans do not provide a more significant number of themes, they still offer unusual combinations. However, Expert AI, which was prompted to be more creative, provided fewer themes than Naive AI, and stayed behind in terms of the unusualness of theme combinations. On the cosine distance, humans provide texts with more semantic distance, meaning that their sequences of words are less predictable, and Expert AI finally outperforms Naive AI. Of note, while \textit{Cosine distance} measures the semantic distance at the individual level (between the sentences of the same text), the measure of \textit{Distance to centroid} represents the group-level semantic distance (between the sentences of one agent). Our results show that humans outperform Naive AI and Expert AI and Naive AI performs better than Expert AI. 

\begin{figure}[h!]
    \centering
    \caption{Text task Embeddings}
    \includegraphics[width=1.1\linewidth]{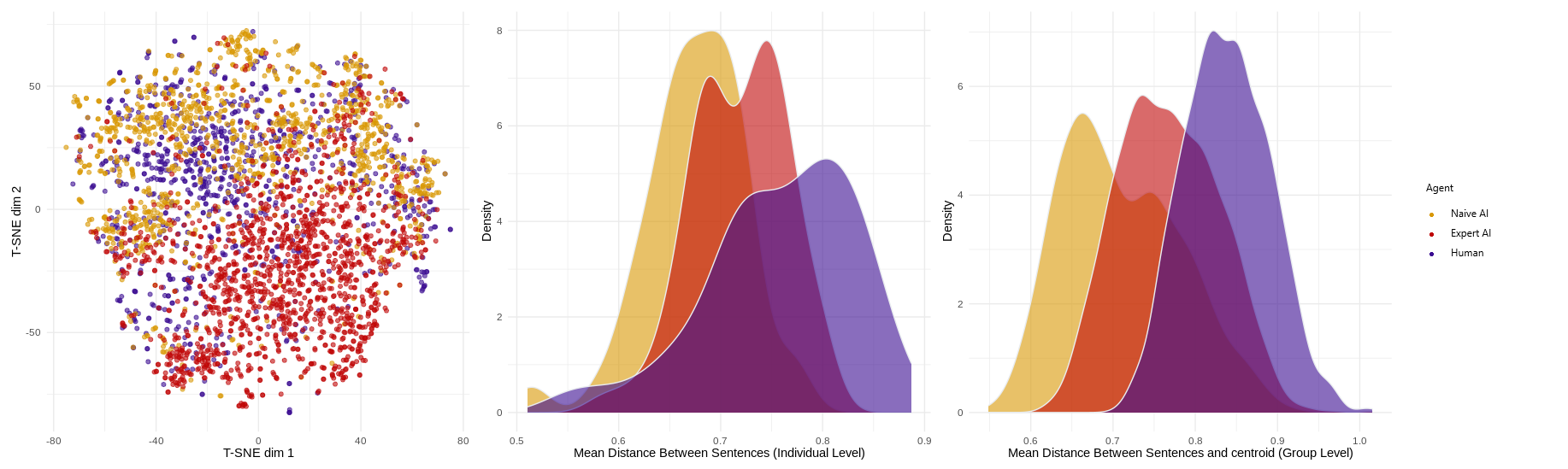}
    \label{fig:text_emb}
\end{figure}

Besides objective measures, we also analysed the creative performance of each agent according to the subjective evaluation provided by human external evaluators. Looking at the comparison between our three agents based on Tukey's HSD test, we find a generalized outperformance of AI over humans on all four creativity criteria. The only differences reside in whether Expert AI also outperforms Naive AI. Table~\ref{tbl:res_summary} presents the results based on our regression presented in Table~\ref{tbl:txt_reg}, in which we use the Naive AI as the reference point and introduce a control on the text length to compare agents' performance. The results show that on \textit{Validity} and \textit{Originality}, Naive AI and humans perform the same while outperforming Expert AI. However, humans fail to provide texts that are as elaborated and high-quality in terms of form or style. Regardless of the criteria, as was the case for subjective measures, Expert AI seems to stay behind.

\begin{table}[!htbp] \centering 
\renewcommand{\arraystretch}{1}
\setlength{\tabcolsep}{0.2pt} 
  \caption{Comparison of Agents' Performance for the Text task}\label{tbl:txt_tukey}
  \label{} 
\scalebox{0.84}{
\begin{threeparttable}
\begin{tabular}{@{\extracolsep{5pt}}lccc} 
\\[-1.8ex]\hline 
\hline \\[-1.8ex] 
Metric & Expert AI vs Naive AI & Human vs Naive AI & Human vs Expert AI \\
\midrule
\addlinespace
Variety & \textcolor{red}{-2.401}$^{***}$ & \textcolor{red}{-2.918}$^{***}$ & -0.517 \\
Theme Frequency (neg.) & 0 & 0.003 & 0.002\\
Uncommonness & \textcolor{red}{-0.162}$^{***}$ & -0.04 & \textcolor{blue}{0.122}$^{***}$\\
Cosine Distance & \textcolor{blue}{0.043}$^{***}$ & \textcolor{blue}{0.083}$^{***}$ & \textcolor{blue}{0.04}$^{***}$ \\
Distance to centroid & \textcolor{blue}{0.059}$^{***}$ & \textcolor{blue}{0.127}$^{***}$ & \textcolor{blue}{0.068}$^{***}$ \\

\addlinespace

Validity & -0.096 & \textcolor{red}{-0.339}$^{***}$ & \textcolor{red}{-0.242}$^{***}$ \\
Form & -0.046 & \textcolor{red}{-0.856}$^{***}$ & \textcolor{red}{-0.81}$^{***}$ \\
Elaboration & \textcolor{blue}{0.191}$^{**}$ & \textcolor{red}{-1.117}$^{***}$ & \textcolor{red}{-1.307}$^{***}$ \\
Originality & \textcolor{blue}{0.221}$^{*}$ & \textcolor{red}{-0.534}$^{***}$ & \textcolor{red}{-0.755}$^{***}$ \\

\addlinespace
\# Words & \textcolor{blue}{106.767}$^{***}$ & \textcolor{red}{-153.27}$^{***}$ & \textcolor{red}{-260.037}$^{***}$ \\
\# Sentences & \textcolor{blue}{6.408}$^{***}$ & \textcolor{red}{-6.139}$^{***}$ & \textcolor{red}{-12.547}$^{***}$ \\
Mean length sentences & \textcolor{red}{-1.449}$^{***}$ & -0.277 & \textcolor{blue}{1.172}$^{***}$ \\
\addlinespace
\hline 
\hline \\[-1.8ex] 
\end{tabular} 
\begin{tablenotes}
\footnotesize
\item \justifying {\it Notes:} This table presents the results of the ANOVA and Tukey's HSD test for the specified variable and group. The coefficients represent the Mean differences between groups. ***, **, and * indicate significance at the 1\%, 5\%, and 10\% levels, respectively.
\end{tablenotes}
 \end{threeparttable}
 }
\end{table}

It is important to note that the relatively low performance of Expert AI is sensitive to the text length control introduced in our analysis. When this control is removed, Expert AI performs better, surpassing both humans and Naive AI in \textit{Elaboration} and \textit{Originality}, while still lagging behind in \textit{Validity} and \textit{Form}. Then, the introduction of this control reveals that Expert AI's initial outperformance was largely driven by the volume of text produced. Once this effect is accounted for, the content does not reflect superior performance. We interpret this as evidence that the sheer quantity of output from AIs may influence human evaluators, leading to higher scores. This control on the length of the texts is also important, as this explains the discrepancy between the results obtained from our regression in Table~\ref{tbl:txt_reg} and the parametric tests run in Table~\ref{tbl:txt_tukey}.

\begin{table}[H] \centering 
\renewcommand{\arraystretch}{0.8}
\setlength{\tabcolsep}{0.05pt} 
  \caption{Polynomial Logit Regression for the Text task} 
  \label{tbl:txt_reg} 
\scalebox{0.84}{
\begin{threeparttable}
\begin{tabular}{@{\extracolsep{5pt}}lcccccccc} 
\\[-1.8ex]\hline 
\hline \\[-1.8ex] 
 & \multicolumn{8}{c}{\textit{Dependent variable:}} \\ 
\cline{2-9} 
\\[-1.8ex] & \multicolumn{2}{c}{Validity} & \multicolumn{2}{c}{Form} & \multicolumn{2}{c}{Elaboration} & \multicolumn{2}{c}{Originality} \\ 
\\[-1.8ex] & (1) & (2) & (3) & (4) & (5) & (6) & (7) & (8)\\ 
\hline \\[-1.8ex] 
\addlinespace
 Expert AI & $-$0.237 & $-$0.606$^{***}$ & $-$0.561$^{***}$ & $-$0.776$^{***}$ & $-$0.533$^{***}$ & $-$0.653$^{***}$ & $-$0.270 & $-$0.368$^{**}$ \\ 
  & (0.182) & (0.207) & (0.173) & (0.187) & (0.175) & (0.191) & (0.170) & (0.177) \\ 
  & & & & & & & & \\ 
 Human & $-$0.495$^{**}$ & $-$0.370 & $-$0.762$^{***}$ & $-$0.766$^{***}$ & $-$0.594$^{***}$ & $-$0.613$^{***}$ & 0.177 & 0.281 \\ 
  & (0.207) & (0.230) & (0.200) & (0.214) & (0.201) & (0.218) & (0.198) & (0.206) \\ 
  & & & & & & & & \\ 
 \# Words & 0.001 & 0.003$^{***}$ & 0.004$^{***}$ & 0.006$^{***}$ & 0.009$^{***}$ & 0.012$^{***}$ & 0.006$^{***}$ & 0.008$^{***}$ \\ 
  & (0.001) & (0.001) & (0.001) & (0.001) & (0.001) & (0.001) & (0.001) & (0.001) \\ 
\vspace{0.5cm}\\
 Evaluator FE  & NO & YES & NO & YES & NO & YES & NO & YES \\ 
  & & & & \\ 
\hline \\[-1.8ex]
AIC & 2257.55 & 1970.32 & 2704.33 & 2405.31 & 2613.2 & 2341.91 & 2985 & 2807 \\ 
Observations & 980 & 980 & 980 & 980 & 980 & 980 & 980 & 980 \\ 
\hline 
\hline \\[-1.8ex] 
\end{tabular} 
\begin{tablenotes}
 \footnotesize
 \item \justifying {\it Notes:} This table presents the coefficients reflecting the impact of agent type on the various creativity criteria scores. ***, ** and * indicate significance at the 1\%, 5\% and 10\% level, respectively. Effects are estimated using an Ordered Polynomial Logit. \end{tablenotes}
 \end{threeparttable}
 }
\end{table} 

\subsection{Alternative Uses task}\label{res:aut}

The Alternative Uses task requires agents to find as many ideas as possible for unusual uses of everyday objects. Of note, all results obtained in Table~\ref{reg:aut} are in line with the parametric tests run in Table~\ref{tbl:aut}. Regarding objective measures, our results show a general outperformance of AI over humans. First, Naive AI surpasses Expert AI in terms of \textit{Variety}, while both outperform humans. The same holds for \textit{Diversity}. However, when considering \textit{Theme frequency} and \textit{Uncommonness}, there was no significant difference between Expert and Naive AI, while both still outperformed humans. While we found no significant effect of \textit{Cosine distance} or \textit{Balance} on our agents' performance, we observe an outperformance of humans over AI agents on group-level semantic distance measured by \textit{Distance to centroid}. These results show that AI provides texts with less frequent themes and more unusual combinations of themes, regardless of the prompting strategy. However, the Naive, which was not prompted to provide more creative texts, still performs the best in the number of themes and the diversity of texts. In any case, human texts never reach the performance of both AI agents except when comparing the semantic distance between agents' pools of ideas. 

Focusing now on subjective measures, Expert and Naive AI clearly exceed humans in all three possible criteria: \textit{Feasibility}, \textit{Originality}, and \textit{Validity}. In other words, AI provides ideas of unusual uses of everyday objects that are perceived as more fitting to the instructions and more implementable while also being more unique. Nonetheless, when comparing our two types of AI agents, Expert AI only performs better than Naive AI in the case of \textit{Originality}.\footnote{This result is the only one differing from the ones obtained in the Tukey's HSD test, where there is no significant difference between Expert and Naive AI.} At the same time, there is no difference between them for \textit{Feasibility} and \textit{Validity}. Here, prompting the AI with a specific requirement to be creative did influence its capacity to provide more original ideas. However, as there is no significant difference between our two prompting strategies for validity and feasibility, our interpretation is that an AI's own perception or definition of creativity is mainly related to originality and not the other criteria. Moreover, as LLM models are thought to provide the most appropriate ideas, it seems logical that both AI agents' performances are similar in feasibility and validity, which both confirm the goodness of fit of the idea to the instructions and the overall task purpose.

Lastly, introducing our control on the average number of words per idea shows a negative effect of longer answers on their originality, with evaluators preferring concision.

\begin{figure}[h!]
    \centering
    \caption{Alternative Uses task Embeddings}
    \includegraphics[width=1.1\linewidth]{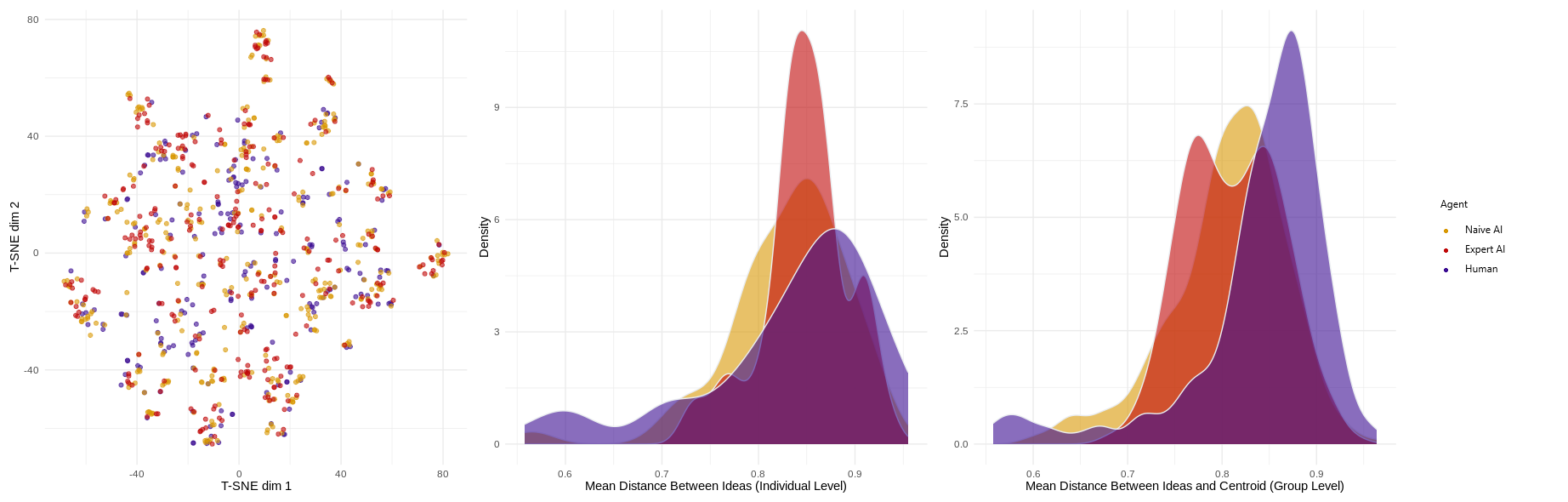}
    \label{fig:aut_emb}
\end{figure}

\begin{table}[h!] 
\centering 
\renewcommand{\arraystretch}{1} 
\setlength{\tabcolsep}{0.2pt} 
\caption{Comparison of Agents' Performance for the Alternative Uses task}\label{tbl:aut}
\label{} 
\scalebox{0.84}{ 
\begin{threeparttable} 
\begin{tabular}{@{\extracolsep{5pt}}lccc} 
\\[-1.8ex]\hline 
\hline \\[-1.8ex] 
Metric & Expert AI vs Naive AI & Human vs Naive AI & Human vs Expert AI\\ 
\midrule 
\addlinespace
Variety & \textcolor{red}{-0.815}$^{***}$ & \textcolor{red}{-2.519}$^{***}$ & \textcolor{red}{-1.704}$^{***}$ \\ 
Balance (neg.)& 0.003 & -0.002 & -0.005 \\ 
Diversity & \textcolor{red}{-0.137}$^{***}$ & \textcolor{red}{-0.636}$^{***}$ & \textcolor{red}{-0.499}$^{***}$ \\ 
Theme Frequency (neg.) & 0 & \textcolor{red}{-0.002}$^{**}$ & \textcolor{red}{-0.003}$^{***}$\\
Uncommonness & -0.139 & \textcolor{red}{-3.657}$^{**}$ & \textcolor{red}{-3.518}$^{**}$\\
Cosine Distance & 0.017 & 0 & -0.017 \\
Distance to centroid & 0.008 & \textcolor{blue}{0.022}$^{***}$ & \textcolor{blue}{0.015}$^{**}$ \\

\addlinespace

Validity & -0.063 & \textcolor{red}{-0.396}$^{***}$ & \textcolor{red}{-0.333}$^{***}$ \\ 
Feasibility  & -0.057 & \textcolor{red}{-0.438}$^{***}$ & \textcolor{red}{-0.38}$^{***}$ \\ 
 
Originality & 0.151 & \textcolor{red}{-0.516}$^{***}$ & \textcolor{red}{-0.667}$^{***}$ \\ 

\addlinespace

Mean \# Words & \textcolor{blue}{1.7}$^{***}$ & \textcolor{red}{-1.295}$^{***}$ & \textcolor{red}{-2.995}$^{***}$ \\
\addlinespace
\hline 
\hline \\[-1.8ex] 
\end{tabular} 
\begin{tablenotes} 
\footnotesize 
\item \justifying {\it Notes:} This table presents the results of the ANOVA and Tukey's HSD test for the specified variable and group. The coefficients represent the Mean differences between groups. ***, **, and * indicate significance at the 1\%, 5\%, and 10\% levels, respectively. 
\end{tablenotes} 
\end{threeparttable} 
} 
\end{table}

\begin{table}[h!] \centering 
\renewcommand{\arraystretch}{0.8}
\setlength{\tabcolsep}{0.05pt} 
  \caption{Polynomial Logit Regression for the Alternative Uses task} 
  \label{reg:aut} 
\scalebox{0.84}{
\begin{threeparttable}
\begin{tabular}{@{\extracolsep{5pt}}lcccccc}  
\\[-1.8ex]\hline 
\hline \\[-1.8ex] 
 & \multicolumn{6}{c}{\textit{Dependent variable:}} \\ 
\cline{2-7} 
\\[-1.8ex] & \multicolumn{2}{c}{Validity} & \multicolumn{2}{c}{Feasibility} & \multicolumn{2}{c}{Originality} \\ 
\\[-1.8ex] & (1) & (2) & (3) & (4) & (5) & (6)\\ 
\hline \\[-1.8ex] 
\addlinespace
Expert AI & 0.079 & $-$0.026 & $-$0.234 & $-$0.302 & 0.584$^{***}$ & 0.817$^{***}$ \\ 
  & (0.220) & (0.265) & (0.262) & (0.282) & (0.222) & (0.241) \\ 
  & & & & & & \\ 
 Human & $-$0.734$^{***}$ & $-$1.330$^{***}$ & $-$1.185$^{***}$ & $-$1.538$^{***}$ & $-$1.078$^{***}$ & $-$1.394$^{***}$ \\ 
  & (0.209) & (0.263) & (0.246) & (0.278) & (0.209) & (0.230) \\ 
  & & & & & & \\ 
 Mean \# Words & $-$0.088$^{**}$ & $-$0.080$^{*}$ & $-$0.081$^{*}$ & $-$0.095$^{**}$ & $-$0.129$^{***}$ & $-$0.169$^{***}$ \\ 
  & (0.039) & (0.048) & (0.043) & (0.048) & (0.042) & (0.046) \\ 
  \vspace{0.5cm}\\
 Evaluator FE  & NO & YES & NO & YES & NO & YES \\ 
  & & & \\ 
\hline \\[-1.8ex] 
AIC & 1253.72 & 942.4 & 959.38 & 905.48 & 1252.88 & 1140.79 \\ 
Observations & 576 & 576 & 576 & 576 & 576 & 576 \\ 
\hline 
\hline \\[-1.8ex] 
\end{tabular} 
\begin{tablenotes}
 \footnotesize
 \item \justifying   {\it Notes:}This table presents the coefficients reflecting the impact of agent type on the various creativity criteria scores. ***, ** and * indicate significance at the 1\%, 5\% and 10\% level, respectively. Effects are estimated using an Ordered Polynomial Logit. \end{tablenotes}
 \end{threeparttable}
 }
\end{table}

\subsection{Draw task}\label{res:img}

Considering the Draw task, the instructions required to draw an alien animal coming from a planet different from Earth. First, we compared our three agents based on Tukey's HSD tests across objective measures and found mixed results considering the possible outperformance of AI over humans. For \textit{Variety} and \textit{Theme frequency}, humans outperform Naive AI. However, there is no significant difference between Expert AI and the two others. The opposite holds for \textit{Uncommonness}, as Naive AI outperforms humans while Expert AI still presents significant differences between the two others. In other words, humans can produce drawings with a larger number of themes and more unique themes, while Naive AI is able to create more uncommon combinations of themes. However, it seems that an AI prompted to be more creative lies in between these two performances with no outperformance of either humans or Naive AI, more as a compromise between them. While no individual level distance measure is computable, we measured group-level \textit{Distance to centroid} and see that human drawings surpass Naive AI, both surpassing Expert AI. 

We also analysed the scores from human evaluators and observed an overall out-performance of Expert AI when generating an original, elaborated, qualitative, and valid drawing. Compared to the two previous tasks, we find a clear-cut between the performance of our two AI agents with a true impact of the prompting strategy pushing the model to be creative. However, focusing on the comparison between Naive AI and Human, the difference is less clear. Considering the drawings \textit{Form} and \textit{Elaboration}, Naive AI outperforms humans. While for \textit{Originality} and \textit{Validity}, there is no significant difference between them. Again, all results obtained in Table~\ref{reg:draw} are in line with the parametric tests run in Table~\ref{tbl:draw}.

It is noteworthy that our control of the complexity of the drawing measured by the share of pixels used in the image is only significant for originality, where a higher number of pixels does contribute positively to the ratings in originality.

\begin{figure}[h!]
    \centering
  \caption{Draw task Embeddings} 
   \includegraphics[width=1.1\linewidth]{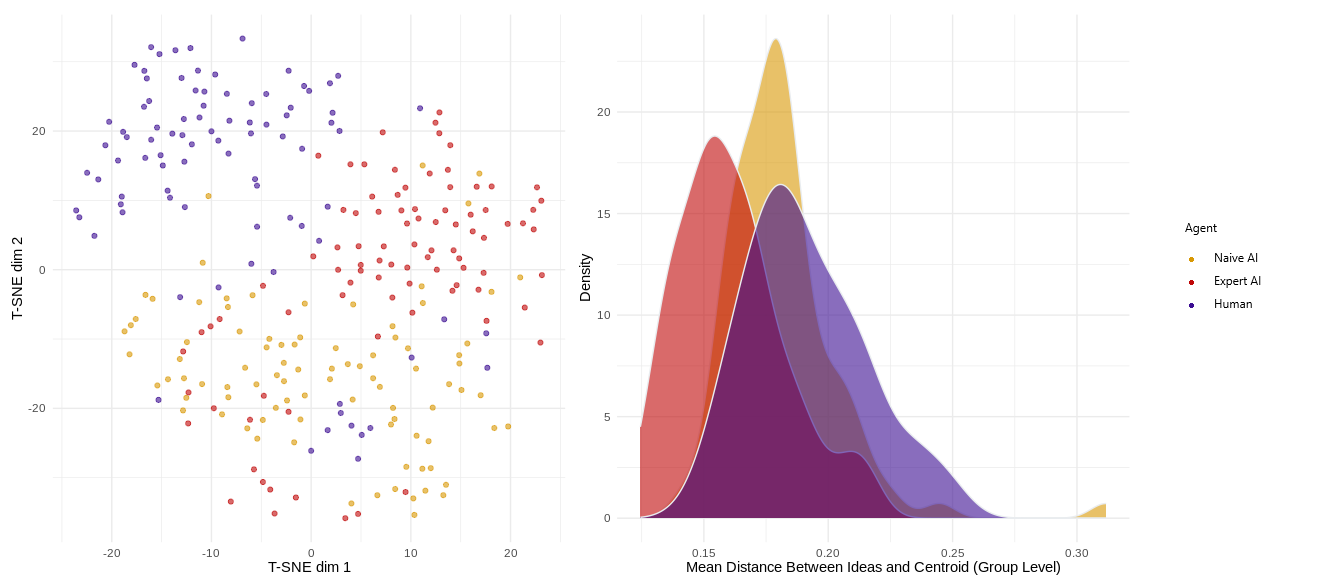}
   \label{fig:draw_emb}
\end{figure}

\begin{table}[!htbp] \centering 
\renewcommand{\arraystretch}{1}
\setlength{\tabcolsep}{0.2pt} 
  \caption{Comparison of Agents' Performance for the Draw task}
  \label{tbl:draw} 
\scalebox{0.84}{
\begin{threeparttable}
\begin{tabular}{@{\extracolsep{5pt}}lccc} 
\\[-1.8ex]\hline 
\hline \\[-1.8ex] 
Metric & Expert AI vs Naive AI & Human vs Naive AI & Human vs Expert AI\\
\midrule
\addlinespace
Variety & 0.733 & \textcolor{blue}{1.222}$^{***}$ & 0.489\\
Theme Frequency (neg.) & 0.002 & \textcolor{blue}{0.008}$^{**}$ & 0.006\\
Uncommonness & -0.012 & \textcolor{red}{-0.033}$^{*}$ & -0.022\\
Distance to centroid & \textcolor{red}{-0.02}$^{***}$ & \textcolor{blue}{0.01}$^{***}$ & \textcolor{blue}{0.03}$^{***}$\\
\addlinespace
Validity & \textcolor{blue}{0.991}$^{***}$ & 0.111 & \textcolor{red}{-0.88}$^{***}$\\
Form & \textcolor{blue}{0.757}$^{***}$ & \textcolor{red}{-0.988}$^{***}$ & \textcolor{red}{-1.745}$^{***}$\\
Elaboration & \textcolor{blue}{0.923}$^{***}$ & \textcolor{red}{-0.769}$^{***}$ & \textcolor{red}{-1.692}$^{***}$\\
Originality & \textcolor{blue}{1.249}$^{***}$ & 0.006 & \textcolor{red}{-1.243}$^{***}$\\

\addlinespace

\% Pixel Used & \textcolor{blue}{0.018}$^{**}$ & \textcolor{red}{-0.143}$^{***}$ & \textcolor{red}{-0.161}$^{***}$\\
\addlinespace
\hline 
\hline \\[-1.8ex] 
\end{tabular} 
\begin{tablenotes}
\footnotesize
\item \justifying {\it Notes:} This table presents the results of the ANOVA and Tukey's HSD test for the specified variable and group. The coefficients represent the Mean differences between groups. ***, **, and * indicate significance at the 1\%, 5\%, and 10\% levels, respectively.
\end{tablenotes}
 \end{threeparttable}
 }
\end{table}

\begin{table}[!htbp] \centering 
\renewcommand{\arraystretch}{0.8}
\setlength{\tabcolsep}{0.05pt} 
  \caption{Polynomial Logit Regression for the Draw task} 
  \label{reg:draw} 
\scalebox{0.84}{
\begin{threeparttable}
\begin{tabular}{@{\extracolsep{5pt}}lcccccccc} 
\\[-1.8ex]\hline 
\hline \\[-1.8ex] 
 & \multicolumn{8}{c}{\textit{Dependent variable:}} \\ 
\cline{2-9} 
\\[-1.8ex] & \multicolumn{2}{c}{Validity} & \multicolumn{2}{c}{Form} & \multicolumn{2}{c}{Elaboration} & \multicolumn{2}{c}{Originality} \\ 
\\[-1.8ex] & (1) & (2) & (3) & (4) & (5) & (6) & (7) & (8)\\ 
\hline \\[-1.8ex] 
\addlinespace
 Expert AI & 1.283$^{***}$ & 1.535$^{***}$ & 1.117$^{***}$ & 1.225$^{***}$ & 1.206$^{***}$ & 1.389$^{***}$ & 1.605$^{***}$ & 1.786$^{***}$ \\ 
  & (0.147) & (0.156) & (0.145) & (0.149) & (0.143) & (0.148) & (0.148) & (0.153) \\ 
  & & & & & & & & \\ 
 Human & 0.086 & 0.105 & $-$1.307$^{***}$ & $-$1.530$^{***}$ & $-$1.101$^{***}$ & $-$1.194$^{***}$ & 0.200 & 0.232 \\ 
  & (0.167) & (0.174) & (0.174) & (0.181) & (0.171) & (0.176) & (0.167) & (0.170) \\ 
  & & & & & & & & \\ 
 \% Pixel Used & $-$0.342 & $-$0.133 & 0.034 & 0.142 & $-$0.425 & $-$0.216 & 1.265$^{*}$ & 1.726$^{**}$ \\ 
  & (0.656) & (0.701) & (0.649) & (0.693) & (0.643) & (0.671) & (0.651) & (0.677) \\ 
\vspace{0.5cm}\\
 Evaluator FE  & NO & YES & NO &  YES & NO & YES & NO &  YES \\ 
  & & & & \\  
\hline \\[-1.8ex] 
AIC & 3105.8 & 2945.98 & 3103 & 3006.52 & 3190.1 & 3114.62 & 3232.35 & 3167.89 \\ 
Observations & 975 & 975 & 975 & 975 & 975 & 975 & 975 & 975 \\ 
\hline 
\hline \\[-1.8ex] 
\end{tabular} 
\begin{tablenotes}
 \footnotesize
 \item \justifying   {\it Notes:}This table presents the coefficients reflecting the impact of agent type on the various creativity criteria scores. ***, ** and * indicate significance at the 1\%, 5\% and 10\% level, respectively. Effects are estimated using an Ordered Polynomial Logit. \end{tablenotes}
 \end{threeparttable}
 }
\end{table} 

\vspace{0.5cm}

\section{Discussion}\label{disc}
 
This study examines the creative performance of Artificial Intelligence (AI hereafter) compared to humans and analyzing the impact of two different prompting strategies on creative outcomes across three different tasks. Based on the central role of creativity in our society and for organisations, the following results provide a deeper understanding of the possible substitution or complementarity of humans and machines when it comes to creative endeavours. In fact, our results give us interesting insights into the creative capacity of models of Generative AI compared to humans. In particular, we observe a superior performance of AI over humans whatever the task. Nevertheless, the distinction between each task remains important as some nuances arise based on the chosen criteria and the objective or subjective nature of the stemming measures. 

When focusing on objective measures, we observe mixed results as our comparison between the creative performance of our three agents depends on the task and the criteria we focus on. For the Text task, there is no outperformance of Expert AI, though it was specifically prompted for better performance. However, when comparing Naive AI and humans, the clearer difference resides in the semantic distance. This can be attributed to the way LMM models operate: they predict the next word based on the current context and probabilities learned from training data. This approach differs significantly from human language generation, which relies on complex neural processes, including synaptic connections and biochemical signals, stemming from emotions, social interactions, and cognitive development over time. Interestingly, the greater distance in human-generated texts does not correlate with better performance in subjective measures such as originality. In fact, as shown in the PCA analysis displayed in Appendix~\ref{app:pca}, this distance captures something distinct and possibly contradictory to subjective measures. Differing from \cite{Koivisto2023}, we controlled for the correlation between our semantic distance measure and originality and found no significant relationship between these two. In other words, the distance observed in human texts does not correlate with more creative outputs according to the subjective criteria assessed by evaluators. This might suggest that the distance in human-generated texts is more likely a result of limited writing skills (in terms of syntax and/or vocabulary etc.) rather than an intentional creative strategy. For the two other tasks, we only see a clear outperformance of AI over humans when it comes to generate ideas of unusual uses of objects. For the Draw task, the most interesting result is the intermediary position of Expert AI that never performs more poorly than Naive AI or humans in all criteria. 

Once we focus on the subjective measures, we can see that the pattern observed previously is not repeated. For the Text task, the evaluators also favoured Naive AI over Expert AI but were ambivalent about the evaluation of human outputs. In fact, for the creative measures specific to the degree of detail and the style of the texts (i.e. elaboration and form), the human performs more poorly than these specific to originality and validity. This points to a duality between the structure and content of creative outputs. In other words, while AI challenges humans in terms of idea form or shape, humans still keep up the pace regarding the ideas' originality and appropriateness. However, we must insist on distinguishing tasks since this observation does not hold for all tasks, especially for the Alternative Use task in which we observe a lower performance of humans for each measure. In addition, for the first time, the Expert AI outperformed the other agents regarding ideas' originality. Finally, for the Draw task, whatever the criterion chosen, the Expert AI performed better than the other two agents, with a positive and significant effect of the prompting strategy pushing the AI model to perform in such a way as to maximize its creative performance. As far as human performance is concerned, the creative outputs are only maintained at the level of these of the Naive AI in terms of originality and validity, similar to the text task, but fails regarding form and elaboration. To summarize, humans appear to outperform AI on content-related criteria (such as originality and validity) but not on formal criteria (such as form and elaboration) only in text-based task, which is the most closed-ended in our study.  

These findings challenge Hypothesis 1 and appear to contradict some previous parts of the literature suggesting that AI excels in closed-ended tasks (e.g.,\cite{Charness2024}). Several factors could explain this discrepancy. First, in \cite{Charness2024}, creativity is measured as a single dimension, with ratings of general creativity on a scale from 0 to 10. In contrast, our study employs a more nuanced measurement of creativity, capturing its multidimensional nature and demonstrating that AI’s advantage over humans is only partial. Then, our findings underscore the importance of the criteria used to evaluate creativity. 

While \cite{Koivisto2023}, which focuses on the AUT, find that AI, on average, outperforms humans except for the best ones, our results show the opposite configuration where AI outperforms humans but is rarely as bad as them. When comparing the distributions of scores for each criterion across tasks and agents, we see that the lowest scores are mainly endowed by humans.\footnote{Comparisons between our three agents per criterion and scores are available in Appendix~\ref{app:comp_top}.} More specifically, for the Text task and the AUT, we observe that for all subjective measures, the proportion of humans with the lowest scores is more important than AI proportions. Including objective measures, we also observe a generally higher variability of scores for humans compared to AI. For the Draw task, the results differ as there are higher proportions of AI in the lower scores but the proportions are still smaller than humans. However, we observe a distinct pattern in the subjective measures: both humans and Naive AI exhibit an inverted U-shaped relationship in score proportions. At the same time, Expert AI shows a consistent increase in proportions, aligning with the direction of the scores. Ultimately, these findings are supported by our measures of semantic distance at the individual and group level, as humans provide texts, ideas, and drawings that are more spread in the semantic space compared to AI agents, which increases the uncertainty or unpredictability of their performance. 

Our findings also challenge Hypothesis 2 regarding close-ended tasks (i.e. Text tasks) as Naive AI outperform Expert AI on subjective and objective measures but distance. One explanation for this surprising result might be related to how Expert AI was designed in our experiment. As explained in Section~\ref{subsec:output_gen}, Expert AI was generated by GPT-4 itself, with no human intervention. It seems that the prompts generated by this AI-based process are excessively focused on the task's constraints, resulting in repetitive and stereotypical outputs featuring knights, princesses, and fairy-tale elements. Paradoxically, instructing GPT-4 to enhance its creative performance led it to overly concentrate on constraints, resulting in poorer scores on some creativity dimensions compared to humans and the naive use of the model. In other words, the Expert model’s rigid adherence to instructions led to a fixation effect, limiting its creativity. Consequently, rather than achieving the goal of an ``Expert AI'' capable of sophisticated and nuanced responses, the experiment revealed that the model was only able to design a ``Meticulous AI'' without effectively leveraging the constraints for creative purposes. This outcome underscores the importance of incorporating human feedback to fully utilize AI's creative potential. Indeed, although we lack direct evidence, it is reasonable to conjecture that a more ``human-in-the-loop'' approach for Expert AI could significantly enhance its performance and potentially surpass human and Naive AI scores.

Once these results have been put into perspective, we can look at their consequences within organisations. Creativity is a central element in the activity of organisations. However, creativity is also risky because of the uncertainty inherent in the process. Therefore, an organisation's rational decision is to focus on the most creative profiles to limit their risk.  Given that our results show that our AI agents outperform human agents and that a comparison of the distribution of their performance by output shows that AI agents tend to perform better even in terms of minimums, it would be natural to see organisations turning away from human ideators in favour of artificial ones. Moreover, we observe a higher uncertainty on the performance of humans based on a higher variability of their performance, which might favour the use of AI in order to minimize the risk for the creative process to fail. However, a second part of our results shows that while AI performs better than humans, the models have certain limitations that require interaction with a human agent to intervene in the prompting strategy. We would like to reiterate an important point concerning the text task. Although Generative AI models are black boxes with millions of parameters whose details we do not know, we do know that they are very sensitive to the way in which a query is conducted. In this study, we chose to let the model optimize this query, and we observed that in the Text task, the model continually created its own fixation effect due to the constraint of using a list of words from the same lexical field. Here, a human agent's intervention would contribute to controlling and avoiding such fixation effects to unlock AI’s full potential. 

To conclude our discussion, it is essential to underline that we controlled for the specific characteristics of our human evaluators to verify if any of their socio-demographic or AI-related information might have influenced their judgment.\footnote{We also accounted for whether the evaluators detected that some outputs were AI-generated or not. Only a few evaluators noticed this, revealing no significant effect on the collected evaluations.} In the end, we could not identify any specific pattern and conclude that evaluators' profiles did not influence the evaluation scores used in our analysis. 

\section{Conclusion}\label{conc}

This study aims to compare the creative performance of AI and humans to identify the strengths and weaknesses of each agent, human or AI. While there is a growing body of work on this subject, this work is the first to examine different tasks based on three levels of task openness and whose performance is assessed by different criteria, integrating two different prompting strategies for the AI agent. To carry out this study, creative outputs were generated by humans and by two distinct types of AI: a Naive AI (which received the same instructions as humans) and an Expert AI (that was prompted to be specifically creative). Once our three agents had generated the outputs, they were submitted for evaluation to external human evaluators who then assessed different creativity criteria for each task; objective measures of output creativity supplemented these subjective measures. In the end, this study on the comparison between the creative capacities of humans and AI allows us to draw lessons both on the study of artificial creativity and the consequences for our society. 

Firstly, further research should consider including more advanced AI evaluation of creativity. This last element, already discussed in the literature \citep{Acar2023}, should also consider which creativity criteria should be computed and whether new ones should be created to cover unexplored or under-explored areas of creativity assessment. In addition, further research should address one limit of our work in considering taking into account creative tasks closer to the real problems organisations face. 

Secondly, our results indicate a clear outperformance of artificial creativity over human creativity. However, they also reveal flaws in Generative AI models in producing creative outputs depending on the prompting strategy adopted. While the first results could lead us to praise the replacement of human creativity by the machine, the second reinforces the idea of human-machine collaboration to maximise creative performance. As already stated for AI technologies in general, ``\textit{to fully exploit the potential of AI, human and machine intelligence must be tightly interwoven}'' \citep[p.19]{Plastino2018}. The use of AI to generate creative outputs is then enriched by human intervention in fine-tuning this AI's prompting to act on the problem-framing, then a central element of the creative process. This necessity of human intervention which becomes all the more important as the constraints intensify.

Finally, it is important to acknowledge that our sample consists exclusively of students, primarily undergraduates, who lack specialized skills in creativity or technical abilities in creative problem-solving. A promising direction for future research would be then to explore whether the observed outperformance of AI holds across specific subpopulations, such as entrepreneurs, managers, engineers, or artists. Investigating both standardized tasks (such as those examined in this study) and sector-specific creative activities could provide valuable insights into the extent of AI's performance relative to human capabilities in various contexts.

\bibliography{references}
\clearpage

\section{Appendix}\label{app}

\subsection{Instructions for data collection and generation}

\subsubsection{Text task}
\vspace{0.3cm}

\textbf{Instructions presented to Human and Naive AI}\footnote{All the following instructions were translated from French to English. The original instructions are available upon request.}: \\

\textit{``In this first part, you are asked to write an interesting and original story using one or more personal memories or experiences with the list of words below. You must use all of the provided words, respecting their singular or plural forms, in addition to any other combination of words of your choice.
The list of words is: 'walls, bricks, towers, roof, keep, stones, rampart, door, window, flag.'
This task is time-limited, and you have a maximum of 15 minutes to complete it.
Once submitted, your text will be evaluated by a jury composed of three other anonymous subjects as follows: your text will be compared with the text of another subject (randomly selected). Each member of the jury will then have to rank the two texts, placing the one they prefer first and the other second.
The subject whose text is ranked first by at least two members of the jury will be awarded a 'jury prize' of €10. The subject whose text is ranked second will win nothing.
The three members of the jury make their decisions in isolation and completely independently (they cannot communicate with each other). Each member must establish a ranking between the two texts. The final ranking is the aggregated ranking of the three jury members. Since the jury is composed of three members, there cannot be a tie between the two texts.
The entire ranking procedure is completely anonymous. You will not know the identity of the jury members who evaluated you, nor the identity of the subject against whom you competed. Similarly, the jury members will not know your identity, nor that of the subject against whom you competed.
Your ranking will be revealed to you at the end of the experiment.''} \\

\noindent \textbf{Prompt generated by GPT to address task instructions for Expert AI} : \\

\textit{``Imagine that you are a storyteller from the Middle Ages, and you must narrate a captivating adventure where an unexpected hero uses elements from his everyday environment—walls, bricks, towers, roof, keep, stones, rampart, door, window, flag—to overcome a series of ingenious challenges. These objects must be central to the plot. You have 15 minutes to weave this story, ensuring it reflects a personal experience or a memorable event from your life, transposed into this medieval universe. Let your imagination run free, remembering that your story will be judged on its originality and ability to captivate an anonymous and independent jury.''}

\subsubsection{Alternative Uses task}
\vspace{0.3cm}

\noindent \textbf{Instructions presented to Human and Naive AI} : \\

\textit{``For this task, we ask you to write down on the provided sheets all the original uses you can think of for a given everyday object. There are certainly common and unoriginal ways to use such an object; for this task, only write down the unusual, creative, and uncommon uses that come to mind.
To help you better understand what is expected, let's take the example of a soda can. Common uses would be ``holding liquid,'' ``serving as a glass,'' ``preserving food,'' etc. However, the uses for which you could receive credit might include ``using it as a flower pot,'' ``a lantern,'' ``a windmill'' (after cutting it), ``a telephone'' (when two are connected by a string), etc.
Here, there will be no constraints in terms of the shape, number, or size of the object.
The task includes five words and will last 15 minutes. You will have three minutes per word to find as many unusual uses as possible. The words will be as follows:
A brick
A cardboard box
An extension cord
A metal pipe
A t-shirt''}\\

\noindent \textbf{Prompt generated by GPT-4 to address task instructions for Expert AI} : \\

\textit{``Immerse yourself in a spirit of unbridled creativity and innovation. For each object on the following list—a brick, a cardboard box, an extension cord, a metal pipe, and a t-shirt—imagine surprising, unorthodox, and inventive uses that transcend their usual functions. Think of applications that would astonish, that would be full of ingenuity, or even poetic or humorous. Let your imagination run wild to transform the ordinary into the extraordinary, revealing unexpected facets of these everyday objects.''}\\

\subsubsection{Draw task}
\vspace{0.3cm}

\noindent \textbf{Instructions presented to Human and Naive AI} : \\

\textit{``For this task, you are asked to imagine and draw an animal from a planet very different from Earth. You will need to draw this animal both from the front and in profile, and to complete your drawing, you will also need to write a short description of the animal and give it a name. For this exercise, you have no constraints other than those mentioned above. You must complete this in a maximum of 15 minutes.''}\\

\noindent \textbf{Prompt generated by GPT-4 to address task instructions for Expert AI} : \\

\textit{``Draw a unique extraterrestrial animal that could exist on a planet very different from Earth. Consider strange and marvellous adaptations that would enable it to survive in unusual environments, such as a thick atmosphere, extreme temperatures, or variable gravity. Your creation should reflect boundless imagination and not be inspired by terrestrial life forms. Draw this animal using only a pencil. The goal is to emphasize the details and unique characteristics of the animal, employing a drawing technique that mimics the style and texture of pencil sketches.''}\\
 
\subsection{Examples of creative outputs}\label{app:examples}

\subsubsection{Text task}
\vspace{0.3cm}
\textbf{\textit{Human}}:\\

\begin{itemize}
    \item \textit{``Je me souviens de cette après-midi où nous avons construit une cabane dans le jardin avec l'aide de ma sœur et de mon père. Mon pauvre père ne savait pas encore dans quelle galère il venait de s'engager. En effet, avec ma sœur, nous aspirions à un château. Nous commencions par choisir un grand bosquet à l'intérieur duquel nous pourrions établir notre château. En guise de murs, nous n'avions pas de briques mais des arbres que mon père a soigneusement taillés pour y insérer des ouvertures qui allaient constituer respectivement une porte d'entrée et une fenêtre. Le feuillage des arbres établissait une belle toiture. Nous avons également laissé pousser un des arbres au milieu de la cabane pour y faire un donjon quelques années plus tard, lorsque l'arbre serait assez grand. Mon père n'en était pas au bout de ses peines, hélas ! Nous l'achevions en lui réclamant un rempart en pierres, avec deux tours, une pour chacune, tout autour de notre château. C'est à la fin de cette longue période de construction que nous organisions une cérémonie d'inauguration de notre magnifique château/cabane, durant laquelle nous plantions notre drapeau juste au-dessus de l'ouverture de la porte. C'était il y a 15 ans. Aujourd'hui, je repense à ces doux moments et me rends compte à quel point l'enfance est une période si courte et si belle... peut-être un peu plus éprouvante pour mon père...''}
    \item \textit{``Voilà ! On y était arrivé. C'était le début de l'après-midi, le mois d'août. Nous étions montés à la vieille ville de Lisbonne. Le soleil brûlait notre peau, fatiguée du trajet. Une partie du rempart de ce beau château était déjà visible, avec ses deux tours, une de chaque côté, un drapeau sur le sommet. On traverse l'entrée. Nous suivons le trajet défini sur le plan qui nous avait été donné. Nous nous trouvons face à un donjon. Nous montons. À l'intérieur, une petite fenêtre nous laisse voir une partie de la ville. Nous sommes finalement arrivés tout en haut du donjon. Et là, cela fut inoubliable. D'un côté la mer, de l'autre cette belle ville qu'on dirait perdue dans le temps. Après avoir profité de ces rayons de soleil et de cette belle vue, on redescend. Il y avait des pierres et des briques partout. Auraient-elles fait partie des murs et de la toiture qui ne sont plus là ? Qui les a construites ? Resteront-elles toujours là ? Nous devons pourtant partir. Quand nous étions sur le point de le faire, mon compagnon aperçoit une petite porte qui n'était pas dans le guide. Nous hésitions. Faut-il aller jeter un coup d'œil ? Enfin, après beaucoup d'hésitation, nous décidons d'y aller. Nous nous approchons de la porte. Nous l'ouvrons. Nous la traversons.''}
    \item \textit{``Il était un temps, à une époque lointaine où le drapeau était utilisé qu'en temps de guerre, les tours n'enfermaient pas de princesses, et le donjon était destiné à héberger le peuple, un monsieur, plus exactement un fabricant de briques faites à partir de pierres précieuses. Ces briques servaient à construire le rempart, la toiture, la fenêtre, les murs, la grande porte d'une grande maison qui un jour allait servir à tout être humain ayant eu un comportement irréprochable. Ce monsieur était un ange.''}
\end{itemize}

\vspace{0.5cm}

\textbf{\textit{Naive AI}}:\\

\begin{itemize}
    \item \textit{``Le soleil se couchait doucement, jetant une lueur dorée sur les murs du vieux château en ruine qui se nichait au sommet de la colline. Les briques, autrefois robustes et imposantes, étaient désormais érodées par le passage du temps, mais elles portaient encore la marque indélébile d'une époque révolue. Les tours dressées vers le ciel, bien qu'affaissées et effondrées par endroits, conservaient leur aura majestueuse. 
    La toiture, jadis ornée d'élégants ardoises, avait cédé sous le poids des siècles, laissant apparaître un ciel parsemé d'étoiles à travers les interstices. L'ombre du donjon se dessinait de manière menaçante dans le lointain, ajoutant une touche mystique à l'ensemble du paysage.
    Je marchais lentement, émerveillé par l'histoire ancienne que racontait chaque pierre du rempart. La porte principale, bien que francs est délabrée, était encore suffisamment robuste pour agir en tant que gardien impénétrable du royaume oublié. La fenêtre à ma droite me donnait un aperçu de l'intérieur du château, où avait autrefois régné l'activité. Tout semblait en paix maintenant, à l'exception de la lueur du soleil couchant qui dansait sur les murs intérieurs, faisant revivre le château pendant un bref instant.
    Un drapeau battait doucement dans la brise au sommet du donjon, déchiré et coloré par le temps, mais il portait encore l'emblème qui avait autrefois été un signe de puissance et de gloire. Le château, avec toutes ses cicatrices, était un symbole frappant de la résilience et de l'endurance, un rappel silencieux que même en ruines, la beauté peut être trouvée.
    Cette promenade dans les ruines du château a été une expérience qui m'a profondément touché. Elle m'a rappelé que même si tout semble perdu, il y a toujours des traces d'espoir et de beauté qui demeurent. Ce château, malgré son état de délabrement, est resté debout, défiant le temps et l'histoire. C'est un symbole d'espoir et de résilience qui continuera à inspirer, même après mon départ.''}
    \item \textit{``C'était un petit village niché au creux d'une vallée. L'héritage de ses glorieux ancêtres se reflétait dans chaque recoin : des murs de briques solides et épais en passant par les tours majestueuses qui surveillaient les montagnes environnantes. Il y avait une toiture en tuiles vernissées qui brillait sous le soleil d'été, et un imposant donjon qui se dressait en plein centre, comme un géant silencieux qui veillait sur ses habitants.
    Des pierres sculptées minutieusement marquaient le chemin vers le rempart, là où résidait le cœur et l'âme du village. C'était un lieu de rassemblement pour les jeunes et les vieux, où les nouvelles étaient échangées, où les enfants couraient librement, les échos de leurs rires résonnant contre les murs de Pierre.
    Au centre du rempart, il y avait une porte imposante, vieille de plusieurs siècles. Ses planches de bois étaient usées, mais elle se tenait toujours fièrement, comme une vieille dame qui avait vu passer les saisons. Elle avait accueilli des rois et des reines, avait été le témoin d'innombrables amours naissantes et de cœurs brisés, et avait survécu à des guerres et des famines.
    Les fenêtres du village étaient comme des yeux qui s'ouvraient sur le monde. Chaque soir, elles s'illuminaient de mille feux, dessinant un tableau vivant de couleurs chaleureuses. Derrière ces fenêtres, chaque habitant du village avait une histoire à raconter, un souvenir à partager.
    Et puis, il y avait le drapeau. Il flottait fièrement au sommet du donjon, agitant ses couleurs vives au gré du vent. Il était un symbole d'unité et de résilience, un rappel que ce village avait survécu à l'épreuve du temps et qu'il continuerait de le faire.
    C'est dans ce village que j'ai grandi, grandissant avec les histoires de mes ancêtres et les rêves d'un futur meilleur. C'est ici que j'ai appris l'importance de la communauté, l'importité de se souvenir de ses racines tout en regardant vers l'avenir. Chaque brique, chaque pierre, chaque tour, chaque morceau de cette toiture, chaque fenêtre, chaque porte, le drapeau, le donjon et le rempart... ils font tous partie de moi, et je fais partie d'eux.''}
    \item \textit{``Laissez-moi vous raconter mon incroyable voyage à travers le temps, dans un château médiéval reculé, encerclé de hauts murs de briques solides. Le lieu avait été témoin de nombreux siècles, et chaque pierre racontait sa propre histoire. Les tours majestueuses du château se dressaient fierement dans le ciel, dominant le paysage environnant avec une grâce sereine.
    La toiture du donjon, d'une architecture exquise, avait miraculeusement résisté au poids des ans. Il y avait quelque chose de véritablement fascinant dans la façon dont le temps semblait s'être arrêté en cette place. Le rempart du château, autrefois un symbole de puissance et de protection, avait subi l'érosion du temps, il était couvert de mousse mais n’en restait pas moins impressionnant.
    Je me suis dirigé vers la porte principale du château. Elle était grande, imposante, forgée en bois massif et renforcée de fer. Autrefois, cette porte devait être le premier rempart de défense contre les envahisseurs. À présent, elle était entrouverte, invitant le voyageur errant à découvrir les secrets qu'elle gardait.
    Je suis entré à l'intérieur, mes pas résonnant sur le sol de pierre. L'intérieur était un mélange d'ombre et de lumière, chaque fenêtre offrant une vue unique sur les terres environnantes. Le château était dépourvu de ses anciens habitants, il restait néanmoins doté d'une vie propre, une vie qu'on pouvait ressentir à chaque coin et recoin.
    Sur le sommet du donjon, flottant au vent, un drapeau déchiré portant le blason du château témoignait d’un passé oublié. Son gracieux ballet, bien qu’usé par le temps, portait en lui l'âme de ce lieu, un dernier hymne à la grandeur d'antan.
    Ce château, à la fois majestueux et mystérieux, était plus qu'un simple tas de pierres. C'était un livre d'histoire vivant, un testament silencieux de l'époque révolue. Et pendant un moment, j'ai eu la chance d'être une petite partie de cette histoire, un simple voyageur traversant les pages du temps.''}
\end{itemize}

\vspace{0.5cm}

\textbf{\textit{Expert AI}}:\\

\begin{itemize}
    \item \textit{``Il était une fois un simple jardinier nommé Fernand, employé du château de Ker-Broc. C'était un homme ordinaire, mais son esprit vivant contrastait avec les murs massifs et les tours intimidantes du château. Chaque jour, il s'occupait des plantes, ramassait les pierres du chemin et admirait le drapeau flottant sur le donjon. Il était émerveillé par les merveilles que sa position humble lui offrait.
    Un jour, le seigneur du château annonça qu'un tournoi aurait lieu, et le gagnant deviendrait son successeur. Fernand fut surpris. Il n'était qu'un jardinier, sans aucune expérience en combat, comment pourrait-il rêver de remporter un tournoi? Pourtant, à chaque fois qu'il passait un rempart ou franchissait la porte grise du château, l'idée lui revenait.
    La veille du tournoi, Fernand ne put fermer l'œil. Il se leva et arpenta les murs en pierre, sophistiqués mais pourtant si familiers pour lui. Dans son esprit, il élabora un plan audacieux. Il connaissait chaque brique et chaque tour du château, chaque fenêtre et chaque fente dans le rempart. Il connaissait les coins et recoins de la toiture du donjon si bien qu'il pourrait s'y déplacer les yeux bandés.
    Le jour du tournoi arriva. Fernand sauta dans l'arène, vêtu de sa tenue de jardinier et armé de son courage. Les rires moqueurs des chevaliers étaient étouffés lorsqu'il utilisa les briques du mur pour déséquilibrer son adversaire. Il grimpa à la tour pour éviter un autre, puis se cacha dans le rempart pour désarçonner le troisième. Il franchit la porte du donjon pour échapper aux autres, utilisa les toitures pour sauter par-dessus eux, puis se précipita vers le drapeau.
    Le public était dans l'incrédulité et émerveillé tandis que Fernand hissait le drapeau à la cime du donjon. Les chevaliers furent stupéfaits par le courage de ce jardinier et le seigneur, impressionné par son ingéniosité, lui donna la victoire. Fernand le jardinier était devenu Fernand le Seigneur.
    Cette histoire fait écho à ma propre expérience. Dans ma vie, j'ai dû subir des défis qui semblaient insurmontables. Mais comme Fernand, j'ai exploité mon environnement familier, mes compétences et mes connaissances pour les surmonter. Nous ne sommes peut-être pas des héros traditionnels, mais nous sommes des héros de notre propre histoire.''}
    \item \textit{``Une fois, dans un ancien royaume, vivait un jeune berger nommé Théo. Loin d'être un héros traditionnel, il passait la majorité de ses journées parmi les moutons et les pierres de l'immense rempart qui protégeait le village. À l'époque, je n'étais qu'un simple enfant, ami de Théo, émerveillé par l'immensité des murs de la cité, qui étaient faites de briques rouges robustes et se prolongeaient jusqu'à des tours menaçantes.
    Un jour, une prophétie s'est abattue sur notre humble village. La prédiction était claire : une énorme créature s'en prendrait à notre donjon et ne laisserait rien derrière elle. Théo, sans aucune compétence de chevalier, a décidé qu'il serait celui qui sauverait le village.
    Sa première épreuve était d'atteindre le donjon. Avec ingéniosité, il a utilisé les briques du rempart pour construire un escalier improvisé jusqu'à la toiture d'une maison. De là, il pourrait atteindre les tours et finalement arriver au donjon.
    En parcourant les murs, il a vu des fenêtres avec des familles terrifiées. Il a vu dans leurs yeux une lueur d'espoir car, pour la première fois, elles voyaient quelqu'un se battre pour elles. Cela l'a motivé encore plus.
    Déjouant la gravité, il a escaladé les tours, se rapprochant de plus en plus du donjon. Son voyage n’a pas été sans dangers. Les pierres du mur se détachaient parfois, rendant son ascension périlleuse. Pourtant, chaque fenêtre qu'il dépassait l'élevait plus haut, l'encourageant à continued.
    Finalement, il a atteint le sommet du donjon, où le drapeau du royaume flottait fièrement. Il l'a remplacé par sa cape de berger, un symbole qu'il était temps pour un simple citoyen de défendre son royaume.
    Alors que la créature approchait, il a utilisé une corde pour descendre à la porte du donjon. Il a pris une pierre et l'a lancée sur la créature. Elle a rugi et a commencé à charger vers le donjon. Avec un sourire, il a fermé la porte du donjon.
    La créature, ne voyant pas l'obstacle, s'est heurtée à la porte et a été assommée. Les villageois ont acclamé Théo, le berger transformé en héros. Son exploit est devenu une histoire racontée encore et encore dans le royaume, en souvenir de l'ingéniosité et du courage d'un berger. Et pour moi, c'est l'histoire d'un ami qui, malgré son manque de force ou de formation, a utilisé les éléments de son environnement quotidien pour devenir le héros que notre village méritait. J'en ai tiré une leçon précieuse : nous sommes tous capables de grandes choses, à condition de faire preuve d'ingéniosité et de courage.''}
    \item \textit{``Il était une fois, dans un village paisible du nom de Cluny, vivait un simple artisan du nom de Francis. Francis, un maçon de son état, vivait sa vie entre les murs de briques qu'il construisait et la douce chaleur de sa maison. Son destin, cependant, était bien plus grand que ce que ces murs pourraient jamais contenir.
    Une nuit, alors que Francis se reposait après une longue journée de labeur, le village fut attaqué par un seigneur féodal connu sous le nom de Baron Noir. Soudain, la quiétude du village s'évanouit, remplacée par le cri des tours de garde, le crépitement des flammes sur les toitures et le grondement des roues du chariot du Baron Noir.
    Le Baron Noir avait un objectif : le donjon de Cluny, où se cachait un trésor mythique. Cependant, le donjon était bien protégé, par des remparts solides et une porte de fer presque impénétrable. Le baron, sûr de sa victoire, installa son drapeau noir au sommet de la colline, en signe de provocation.
    Francis, de son humble demeure, observait la vue par la fenêtre. Avec chaque pierre de son village qui tombait, son cœur se serrait de douleur. Il ne pouvait pas rester les bras croisés. Il devait agir.
    Armé de son expertise, il élabora un plan. Avec les mêmes briques qu'il utilisait pour construire de simples murs, Francis commença à construire des pièges autour du donjon. Il renforça les remparts, ajouta des blocs de pierres pour créer un labyrinthe, installa une fausse porte pour tromper l'ennemi.
    La nuit tomba et le Baron Noir avança, sûr de sa victoire. Mais à chaque tour, à chaque recoin, il était confronté à un nouveau défi. Les murs qui semblaient fragiles était presque impénétrables, les remparts étaient insurmontables, et la porte qu'il pensait être l'entrée du donjon n'était rien d'autre qu'un piège.
    L'aube se leva et le Baron Noir n'était nulle part en vue. Le drapeau noir avait été remplacé par celui du village. Francis, de sa fenêtre, souriait en regardant son œuvre. Il avait réussi. Il avait utilisé les éléments les plus ordinaires de son quotidien pour protéger son village, pour être un héros.
    Cette histoire, mes amis, me rappelle ma propre expérience. Tout comme Francis, j'ai dû utiliser les outils de mon quotidien pour surmonter les défis qui se présentaient à moi. C'est une leçon que nous devrions tous retenir : peu importe combien une tâche peut paraître grande, avec de l'ingéniosité et de la détermination, rien n'est impossible.''}
\end{itemize}

\subsubsection{Alternative Uses task}

\vspace{0.5cm}

\textbf{\textit{Human}}:\\

\begin{itemize}
    \item Brique: \textit{une arme de jet, une haltère,	une marche pour faire du step, un chapeau,	un instrument de musique percussion (adapté pour un public souffrant d'hyperacousie)}.
    \item Boîte en carton: \textit{jeu pour enfants ou chats,	support peinture,	décoration (après découpage),	luge}.
\end{itemize}

\vspace{0.5cm}

\noindent \textbf{\textit{Naive AI}}:\\

\begin{itemize}
    \item Brique: \textit{presser des vêtements en remplacement d'un fer à repasser,	outil de cuisson pour le poulet rôti à la brique,	support de livre,	altère de fortune pour faire de l'exercice,	bloc de refroidissement pour les ordinateurs portables,	instrument de musique à percussion	sabot de frein pour les vélos}.	
    \item Boîte en carton:	\textit{maison pour chat ou cabane,	décor de théâtres pour enfants,	pot de plantation biodegradable,	support d’écran pour créer un cinéma maison avec un smartphone,	mur d'isolation phonique	abat-jour pour lampe,	chaise ou tabouret (si assez robuste),	costume de robot pour Hallowen}.	
\end{itemize}

\vspace{0.5cm}

\noindent \textbf{\textit{Expert AI}}:\\

\begin{itemize}
    \item Brique:	\textit{porte-livre sur une étagère,	presse-papier de bureau,	support de yoga pour des poses de flexibilité,	four solaire miniature en la couvrant de feuille d'aluminium,	moule à glace pour faire des glaçons géants,	une percussion musicale en la frappant avec un marteau dans un concert de musique expérimentale}.
    \item Boîte en carton:	\textit{un théâtre d'ombre miniature pour des séances de conte nocturnes,	une maison pour jouer pour les enfants,	un espace de rangement secret sous le lit,	un attrape-rêves décoratif une fois découpée et peinte.	un bateau de marin d'eau douce dans une mare,	un support pour des semis dans un potager urbain,	une cabine à peindre pour libérer sa créativité artistique}.	
\end{itemize}

\subsubsection{Draw task}

\vspace{1cm}
\begin{center}
    \includegraphics[scale=0.175]{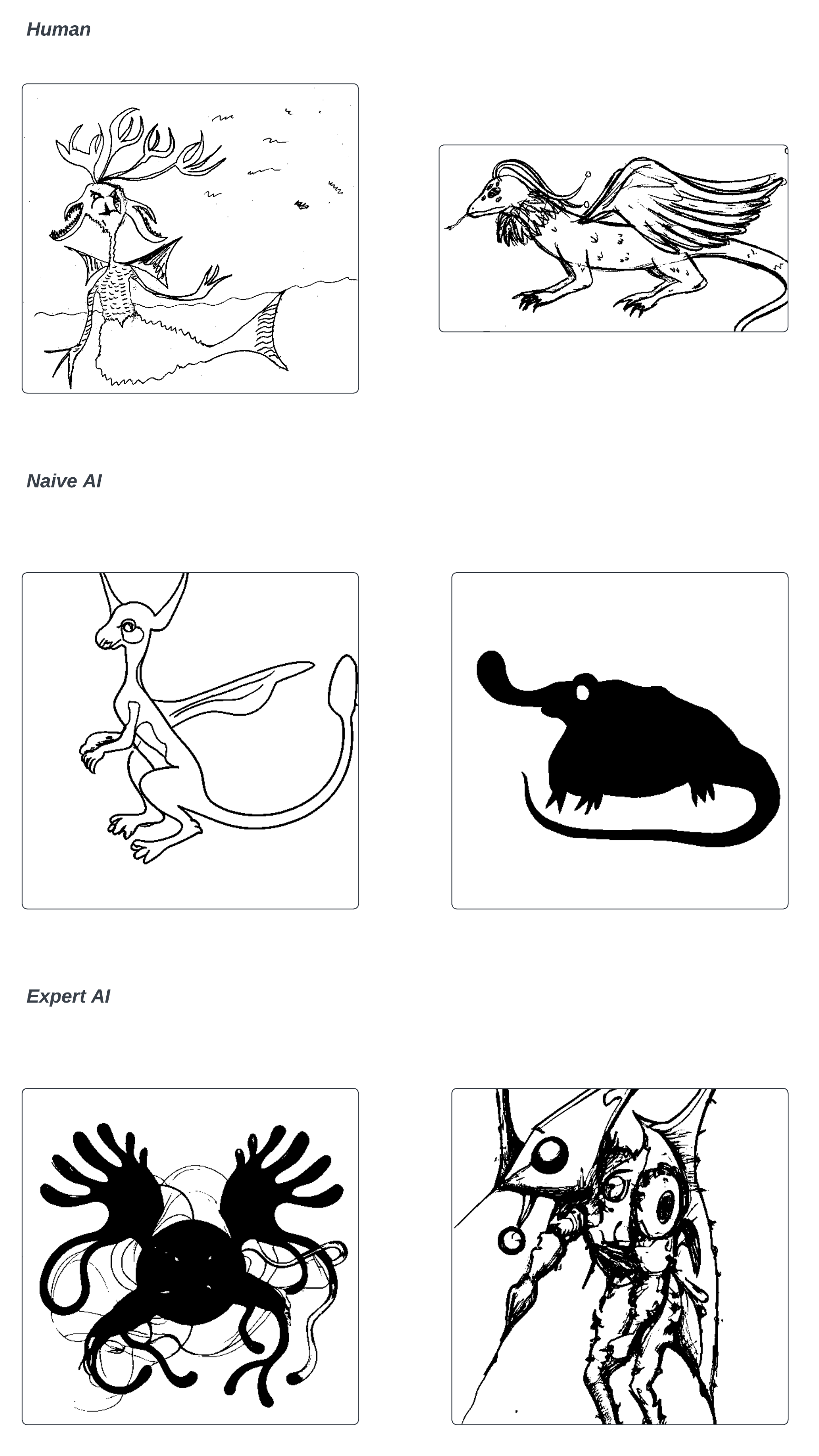}
\end{center}

\subsection{Instructions for creative assessment}\label{app:exp_instr}

\subsubsection{General instructions}
\vspace{0.3cm}

\textit{``Welcome to this social science experiment. We sincerely thank you for your participation. Before we begin, we would like to remind you of the importance of responding seriously and thoroughly to the questions posed so that this experiment can yield valid and relevant results.
All your responses will be collected anonymously.
In this experiment, you will perform the role of an evaluator of various texts. You will be asked to assess a series of 15 texts according to different criteria.
At the end of the experiment, you will also need to complete a questionnaire.
For your participation in the entire experiment, you will be compensated 15€.''}

\subsubsection{Text task}
\vspace{0.3cm}

\textit{``For this task, a series of texts will appear on the screen one after the other, randomly. The texts have been created following the instructions below:}

\vspace{0.3cm}

\noindent \textit{Instructions: You are asked to write an interesting and original story using one or more personal memories or experiences with the list of words below. You must use all of the proposed words, respecting their singular or plural forms, in addition to any other combinations of words of your choice. The list of words is “walls, bricks, towers, roof, keep, stones, rampart, door, window, flag.”}

\vspace{0.3cm}

\noindent \textit{For each displayed text, you are asked to rate the text based on an evaluation grid consisting of four criteria:
\vspace{0.3cm}
\begin{itemize}
    \item Validity: adherence to instructions 0 to 5
    \item Form: writing style, syntax 0 to 5
    \item Elaboration: level of detail/description of the story 0 to 5
    \item Originality: 0 to 5
\end{itemize}
\vspace{0.3cm}
A total of 15 texts will be displayed on the screen.}

\subsubsection{Alternative Uses task}
\vspace{0.3cm}

\textit{``For this task, several lists of ideas for unusual uses of a common object will be displayed successively and randomly on the screen. The lists have been created following the instructions below:}

\vspace{0.3cm}

\noindent \textit{Instructions: For this task, we ask you to write down on the sheets provided all the original uses of a given everyday object. There are certainly common and unoriginal ways to use such an object; for this task, only write down the unusual, creative, and uncommon uses that you can think of.
To help you better understand what is expected, let’s take the example of a soda can. Common uses would be “containing liquid,” “serving as a glass,” “preserving food,” etc. Whereas uses for which you could be credited might be: “using it as a flower pot,” “a lantern,” “a windmill” (after cutting), “a phone” (when two are connected by a wire), etc.
There will be no constraints regarding the shape, number, or size of the object.}

\vspace{0.3cm}

\noindent \textit{For each list displayed, you are asked to rate each idea using a scoring grid consisting of three criteria:\\
\vspace{0.3cm}
\begin{itemize}
    \item Validity: adherence to the instructions 0 to 5
    \item Feasibility: the suggested use of the object is doable/viable 0 to 5
    \item Originality: 0 to 5
\end{itemize}
\vspace{0.3cm}
A total of 15 lists of ideas will be displayed on the screen.''}

\subsubsection{Draw task}
\vspace{0.3cm}

\textit{``For this task, a series of drawings will be displayed successively and randomly on the screen. The drawings were created following the instructions below:}

\vspace{0.3cm}

\noindent \textit{Instructions: For this task, you are asked to imagine and draw an animal from a planet very different from Earth.}

\vspace{0.3cm}

\noindent \textit{For each drawing displayed, you are asked to rate the drawing using an evaluation grid consisting of four criteria:
\vspace{0.3cm}
\begin{itemize}
\item Validity: adherence to instructions 0 to 5
\item Form: style of the drawing, graphic quality 0 to 5
\item Elaboration: level of detail/precision of the drawing 0 to 5
\item Originality 0 to 5
\end{itemize}
\vspace{0.3cm}\
A total of 15 drawings will be displayed on the screen.''}

\vspace{0.5cm}

\subsection{Questionnaires}\label{app:quest}

\subsubsection{Socio-demographic questionnaire}
\vspace{0.5cm}

\begin{multicols}{2}
\begin{enumerate}
    \item Your age: \_\_\_\_\_\_
\vspace{0.25cm}
\item Your gender: \\
\makebox[1.5cm][l]{$\square$}Male \\
\makebox[1.5cm][l]{$\square$}Female

\item The type of degree you are enrolled in: \\
\makebox[1.5cm][l]{$\square$}Bachelor's \\
\makebox[1.5cm][l]{$\square$}Master's \\
\makebox[1.5cm][l]{$\square$}Doctorate \\
\makebox[1.5cm][l]{$\square$}Other (specify) \_\_\_\_\_\_
\vspace{0.25cm}
\item Your field of study: \\
\makebox[1.5cm][l]{$\square$}Law Economics-Management \\
\makebox[1.5cm][l]{$\square$}Literature-Languages \\
\makebox[1.5cm][l]{$\square$}Exact Sciences \\
\makebox[1.5cm][l]{$\square$}Psychology-Sociology \\
\makebox[1.5cm][l]{$\square$}Political Science \\
\makebox[1.5cm][l]{$\square$}Other (specify) \_\_\_\_\_\_
\vspace{0.25cm}
\item Is French your native language? \\
\makebox[1.5cm][l]{$\square$}Yes \\
\makebox[1.5cm][l]{$\square$}No (If no, which one?) \_\_\_\_\_\_
\vspace{0.25cm}
\item Among the following hobbies, check those you have or are currently practising: \\
\makebox[1.5cm][l]{$\square$}Writing \\
\makebox[1.5cm][l]{$\square$}Drawing \\
\makebox[1.5cm][l]{$\square$}Painting \\
\makebox[1.5cm][l]{$\square$}Reading \\
\makebox[1.5cm][l]{$\square$}Visual Arts \\
\makebox[1.5cm][l]{$\square$}Cinema/Theater
\vspace{0.25cm}
\item On a scale of 1 to 10, how would you rate your proficiency in the French language (spelling, style, syntax, etc.)?
\end{enumerate}
\end{multicols}

\subsubsection{AI questionnaire}
\vspace{0.5cm}

\begin{multicols}{2}
\begin{enumerate}
\item During this experiment, we presented you with different productions to evaluate. These productions were created in two distinct ways. In your opinion, what are these two ways and how did you identify them? (If you do not know how to answer this question, click ``Next'') \
\vspace{0.25cm}
\item During this experiment, we presented you with different productions to evaluate. Half of these productions were generated by humans and the other half by AI. (Yes or No)\
\vspace{0.25cm}
\item If yes, did this influence your judgment? \\
\begin{itemize}
    \item Yes, I think I judged the AI-generated productions more favorably
    \item Yes, I think I judged the human-generated productions more favorably 
    \item It did not impact my judgment
\end{itemize}
\vspace{0.25cm}
\item How would you rate your knowledge of artificial intelligence? (1 being ``I know absolutely nothing'' and 5 being ``I have advanced knowledge'')\
\vspace{0.25cm}
\item In your daily life (outside of work or studies), do you use artificial intelligence tools? \\
\makebox[1.5cm][l]{$\square$} Never \\
\makebox[1.5cm][l]{$\square$} Rarely \\
\makebox[1.5cm][l]{$\square$} Often \\
\makebox[1.5cm][l]{$\square$} Regularly\
\vspace{0.25cm}
\item Which applications? \\
\makebox[1.5cm][l]{$\square$} GPT-4, Bing, Bard \\
\makebox[1.5cm][l]{$\square$} Midjourney, DALL-E \\
\makebox[1.5cm][l]{$\square$} DeepL \\
\makebox[1.5cm][l]{$\square$} Others: \
\vspace{0.25cm}
\item In your work or studies, do you use artificial intelligence tools? \\
\makebox[1.5cm][l]{$\square$} Never \\
\makebox[1.5cm][l]{$\square$} Rarely \\
\makebox[1.5cm][l]{$\square$} Often \\
\makebox[1.5cm][l]{$\square$} Regularly\
\vspace{0.25cm}
\item Which applications? \\
\makebox[1.5cm][l]{$\square$} GPT-4, Bing, Bard \\
\makebox[1.5cm][l]{$\square$} Midjourney, DALL-E \\
\makebox[1.5cm][l]{$\square$} DeepL \\
\makebox[1.5cm][l]{$\square$} Others: \
\vspace{0.25cm}
\item AI applications are a good way to improve what I create. (From 1 ``Strongly disagree'' to 5 ``Strongly agree''.)\
\vspace{0.25cm}
\item AI applications are a good way to generate relevant content. (From 1 ``Strongly disagree'' to 5 ``Strongly agree''.)\
\vspace{0.25cm}
\item How would you rate your level of creativity? (From 1 to 5.)\
\vspace{0.25cm}
\item Do you think an AI can be creative? (Yes or No)\
\end{enumerate}
\end{multicols}

\vspace{0.5cm}

\newpage

\subsubsection{Descriptive statistics for each task}
\vspace{0.3cm}
\subsubsection*{Text task}
\begin{table}[H] \centering 
\renewcommand{\arraystretch}{1}
\setlength{\tabcolsep}{0.2pt} 
  \caption{Text task Descriptive Statistics}\label{}
  \label{} 
\scalebox{0.84}{
\begin{threeparttable}
\begin{tabular}{@{\extracolsep{5pt}}lccc} 
\\[-1.8ex]\hline 
\hline \\[-1.8ex] 
Variable & Naive AI & Expert AI & Human\\
\midrule
\addlinespace
Mean Validity & 4.418 (0.818) & 4.322 (0.889) & 4.080 (1.074)\\
Mean Form & 4.012 (0.913) & 3.966 (0.886) & 3.156 (1.31)\\
Mean Elaboration & 3.945 (0.856) & 4.136 (0.849) & 2.829 (1.311)\\
Mean Originality & 3.506 (1.103) & 3.728 (1.063) & 2.972 (1.445)\\
\# Evaluations & 330 & 323 & 327\\
\addlinespace
\midrule
\addlinespace
Mean \# Words & 380.424 (46.22) & 486.735 (51.071) & 226.845 (89.011)\\
Mean \# Sentences & 17.506 (3.383) & 23.940 (3.782) & 11.381 (5.346)\\
Mean Mean length sentences & 22.106 (2.58) & 20.612 (2.376) & 21.737 (7.126)\\
Mean Cosine Distance & 0.676 (0.049) & 0.719 (0.046) & 0.759 (0.072)\\
Mean Variety & 7.847 (2.146) & 5.446 (1.484) & 4.929 (1.678)\\
Mean Theme Frequency (neg.) & -0.030 (0.018) & -0.030 (0.019) & -0.028 (0.019)\\
Mean Uncommonness & -0.310 (0.116) & -0.472 (0.194) & -0.350 (0.178)\\
\# Observations & 85 & 83 & 84\\
\addlinespace
\hline 
\hline \\[-1.8ex] 
\end{tabular} 
 \end{threeparttable}
 }
\end{table} 

\subsubsection*{Alternative Uses task}
\begin{table}[H] \centering 
\renewcommand{\arraystretch}{1}
\setlength{\tabcolsep}{0.2pt} 
  \caption{Alternative Uses task Descriptive Statistics}
  \label{} 
\scalebox{0.84}{
\begin{threeparttable}
\begin{tabular}{@{\extracolsep{5pt}}lccc} 
\\[-1.8ex]\hline 
\hline \\[-1.8ex] 
Variable & Naive AI & Expert AI & Human\\
\midrule
\addlinespace
Mean Validity & 4.495 (0.85) & 4.432 (0.947) & 4.099 (1.272)\\

Mean Feasibility  & 4.729 (0.686) & 4.672 (0.657) & 4.292 (1.152)\\
Mean Originality  & 4.427 (0.841) & 4.578 (0.697) & 3.911 (1.24)\\
\# Evaluations & 192 & 192 & 192\\
\addlinespace
\midrule
\addlinespace
Mean \# word mean & 6.345 (1.753) & 8.073 (1.735) & 5.022 (2.585)\\
Mean Cosine Distance & 0.831 (0.064) & 0.848 (0.045) & 0.831 (0.096)\\
Mean Variety & 5.796 (1.155) & 4.981 (0.789) & 3.278 (1.393)\\
Mean Balance (neg.)& 0.952 (0.047) & 0.955 (0.053) & 0.950 (0.075)\\
Mean Diversity & 1.698 (0.231) & 1.561 (0.173) & 1.062 (0.468)\\
Mean Theme Frequency (neg.)& -0.030 (0.018) & -0.030 (0.019) & -0.028 (0.019)\\
Mean Uncommonness & -0.310 (0.116) & -0.472 (0.194) & -0.350 (0.178)\\
\# Observations & 54 & 54 & 54\\
\addlinespace
\hline 
\hline \\[-1.8ex] 
\end{tabular} 
 \end{threeparttable}
 }
\end{table} 

\subsubsection*{Draw task}
\begin{table}[H] \centering 
\renewcommand{\arraystretch}{1}
\setlength{\tabcolsep}{0.2pt} 
  \caption{Draw task Descriptive Statistics}
  \label{} 
\scalebox{0.84}{
\begin{threeparttable}
\begin{tabular}{@{\extracolsep{5pt}}lccc} 
\\[-1.8ex]\hline 
\hline \\[-1.8ex] 
Variable & Naive AI  & Expert AI & Human\\
\midrule
\addlinespace
Mean Validity & 3.071 (1.559) & 4.062 (1.203) & 3.182 (1.54)\\
Mean Form & 3.031 (1.303) & 3.788 (1.218) & 2.043 (1.319)\\
Mean Elaboration & 2.692 (1.355) & 3.615 (1.287) & 1.923 (1.387)\\
Mean Originality & 2.548 (1.449) & 3.797 (1.268) & 2.554 (1.516)\\
\# Evaluations & 325 & 325 & 325\\
\addlinespace
\midrule
\addlinespace
Mean \% Pixel Used & 0.221 (0.115) & 0.238 (0.095) & 0.075 (0.042)\\
Mean Variety & 6.811 (2.365) & 7.544 (2.558) & 8.033 (2.524)\\
Mean Theme frequency & -0.031 (0.022) & -0.029 (0.024) & -0.023 (0.017)\\
Mean Uncommonness & -0.704 (0.103) & -0.716 (0.094) & -0.737 (0.09)\\
\# Observations & 90 & 90 & 90\\
\addlinespace
\hline 
\hline \\[-1.8ex] 
\end{tabular} 
 \end{threeparttable}
 }
\end{table} 

\vspace{0.5cm}

\subsubsection{Pairwise Wilcoxon Test}
\vspace{0.3cm}
\subsubsection*{Text task}
\begin{table}[!htbp] \centering 
\renewcommand{\arraystretch}{1}
\setlength{\tabcolsep}{0.2pt} 
  \caption{Comparison of Agents’ Performance for the Text task (non-parametric)}
  \label{} 
\scalebox{0.84}{
\begin{threeparttable}
\begin{tabular}{@{\extracolsep{5pt}}lccc} 
\\[-1.8ex]\hline 
\hline \\[-1.8ex] 
Metric & Expert AI vs Naive AI & Human vs Naive AI & Human vs Expert AI\\
\midrule
Variety & \textcolor{red}{-2.401}$^{***}$ & \textcolor{red}{-2.918}$^{***}$ & \textcolor{red}{-0.517}$^{**}$\\
Theme Frequency (neg.) & 0 & 0.003 & 0.002\\
Uncommonness & \textcolor{red}{-0.162}$^{***}$ & -0.04 & \textcolor{blue}{0.122}$^{***}$\\
Cosine distance & \textcolor{blue}{0.043}$^{***}$ & \textcolor{blue}{0.083}$^{***}$ & \textcolor{blue}{0.04}$^{***}$ \\
Distance to centroid & \textcolor{blue}{0.059}$^{***}$ & \textcolor{blue}{0.127}$^{***}$ & \textcolor{blue}{0.068}$^{***}$\\
\addlinespace
Validity & -0.096 & \textcolor{red}{-0.339}$^{***}$ & \textcolor{red}{-0.242}$^{**}$ \\
Form & -0.046 & \textcolor{red}{-0.856}$^{***}$ & \textcolor{red}{-0.81}$^{***}$ \\
Elaboration & \textcolor{blue}{0.191}$^{***}$ & \textcolor{red}{-1.117}$^{***}$ & \textcolor{red}{-1.307}$^{***}$ \\
Originality & \textcolor{blue}{0.221}$^{**}$ & \textcolor{red}{-0.534}$^{***}$ & \textcolor{red}{-0.755}$^{***}$ \\
\addlinespace
\# Words & \textcolor{blue}{106.767}$^{***}$ & \textcolor{red}{-153.27}$^{***}$ & \textcolor{red}{-260.037}$^{***}$ \\
\# Sentences & \textcolor{blue}{6.408}$^{***}$ & \textcolor{red}{-6.139}$^{***}$ & \textcolor{red}{-12.547}$^{***}$ \\
Mean length sentences & \textcolor{red}{-1.449}$^{***}$ & \textcolor{red}{-0.277}$^{***}$ & 1.172 \\
\addlinespace
\hline 
\hline \\[-1.8ex] 
\end{tabular} 
\begin{tablenotes}
\footnotesize
\item \justifying {\it Notes:} This table presents the results of the Pairwise Wilcoxon test for the specified variable and group. The coefficients represent the Mean differences between groups. ***, **, and * indicate significance at the 1\%, 5\%, and 10\% levels, respectively.
\end{tablenotes}
 \end{threeparttable}
 }
\end{table}

\subsubsection*{Alternative Uses task}
\begin{table}[H] 
\centering 
\renewcommand{\arraystretch}{1} 
\setlength{\tabcolsep}{0.2pt} 
\caption{Comparison of Agents’ Performance for the Alternative Uses task (non-parametric)} 
\label{} 
\scalebox{0.84}{ 
\begin{threeparttable} 
\begin{tabular}{@{\extracolsep{5pt}}lccc} 
\\[-1.8ex]\hline 
\hline \\[-1.8ex]  
Metric & Expert AI vs Naive AI & Human vs Naive AI & Human vs Expert AI\\ 
\midrule 
\addlinespace
Variety & \textcolor{red}{-0.815}$^{***}$ & \textcolor{red}{-2.519}$^{***}$ & \textcolor{red}{-1.704}$^{***}$\\
Balance (neg.) & 0.003 & -0.002 & -0.005\\
Diversity & \textcolor{red}{-0.137}$^{***}$ & \textcolor{red}{-0.636}$^{***}$ & \textcolor{red}{-0.499}$^{***}$\\
Theme Frequency (neg.) & 0 & -0.002 & \textcolor{red}{-0.003}$^{**}$\\
Uncommonness & -0.139 & \textcolor{red}{-3.657}$^{**}$ & \textcolor{red}{-3.518}$^{*}$\\
Distance to centroid & 0.008 & \textcolor{blue}{0.022}$^{***}$ & \textcolor{blue}{0.015}$^{***}$\\
\addlinespace
Validity & -0.062 & \textcolor{red}{-0.396}$^{***}$ & \textcolor{red}{-0.333}$^{**}$ \\ 
Feasibility  & -0.057 & \textcolor{red}{-0.438}$^{***}$ & \textcolor{red}{-0.38}$^{***}$ \\ 
Originality & 0.151 & \textcolor{red}{-0.516}$^{***}$ & \textcolor{red}{-0.667}$^{***}$\\
\addlinespace
Mean \# Words & \textcolor{blue}{1.7}$^{***}$ & \textcolor{red}{-1.295}$^{***}$ & \textcolor{red}{-2.995}$^{***}$ \\ 
\addlinespace
\hline 
\hline \\[-1.8ex] 
\end{tabular} 
\begin{tablenotes} 
\footnotesize 
\item \justifying {\it Notes:} This table presents the results of the Pairwise Wilcoxon test for the specified variable and group. The coefficients represent the Mean differences between groups. ***, **, and * indicate significance at the 1\%, 5\%, and 10\% levels, respectively. 
\end{tablenotes} 
\end{threeparttable} 
} 
\end{table}

\subsubsection*{Draw task}
\begin{table}[!htbp] \centering 
\renewcommand{\arraystretch}{1}
\setlength{\tabcolsep}{0.2pt} 
  \caption{Comparison of Agents’ Performance for the Draw task (non-parametric)}
  \label{} 
\scalebox{0.84}{
\begin{threeparttable}
\begin{tabular}{@{\extracolsep{5pt}}lccc} 
\\[-1.8ex]\hline 
\hline \\[-1.8ex] 
Metric & Expert AI vs Naive AI & Human vs Naive AI & Human vs Expert AI\\
\midrule
Variety & 0.733 & \textcolor{blue}{1.222}$^{***}$ & 0.489\\
Theme Frequency (neg.) & 0.002 & 0.008 & 0.006\\
Uncommonness & -0.012 & \textcolor{red}{-0.033}$^{**}$ & -0.022\\
Distance to centroid & \textcolor{red}{-0.02}$^{***}$ & \textcolor{blue}{0.01}$^{***}$ & \textcolor{blue}{0.03}$^{***}$ \\
\addlinespace
Validity & \textcolor{blue}{0.991}$^{***}$ & 0.111 & \textcolor{red}{-0.88}$^{***}$\\
Form & \textcolor{blue}{0.757}$^{***}$ & \textcolor{red}{-0.988}$^{***}$ & \textcolor{red}{-1.745}$^{***}$\\
Elaboration & \textcolor{blue}{0.923}$^{***}$ & \textcolor{red}{-0.769}$^{***}$ & \textcolor{red}{-1.692}$^{***}$\\
Originality & \textcolor{blue}{1.249}$^{***}$ & 0.006 & \textcolor{red}{-1.243}$^{***}$\\

\addlinespace
\% Pixel Used & \textcolor{blue}{0.018}$^{***}$ & \textcolor{red}{-0.143}$^{***}$ & \textcolor{red}{-0.161}$^{***}$\\
\addlinespace
\hline 
\hline \\[-1.8ex] 
\end{tabular} 
\begin{tablenotes}
\footnotesize
\item \justifying {\it Notes:} This table presents the results of the Pairwise Wilcoxon test for the specified variable and group. The coefficients represent the Mean differences between groups. ***, **, and * indicate significance at the 1\%, 5\%, and 10\% levels, respectively.
\end{tablenotes}
 \end{threeparttable}
 }
\end{table}

\newpage

\subsection{PCA Analysis}\label{app:pca}
\subsubsection{Text task}
\begin{figure}[h!]
    \centering
    \caption{}
    \includegraphics[width=0.75\linewidth]{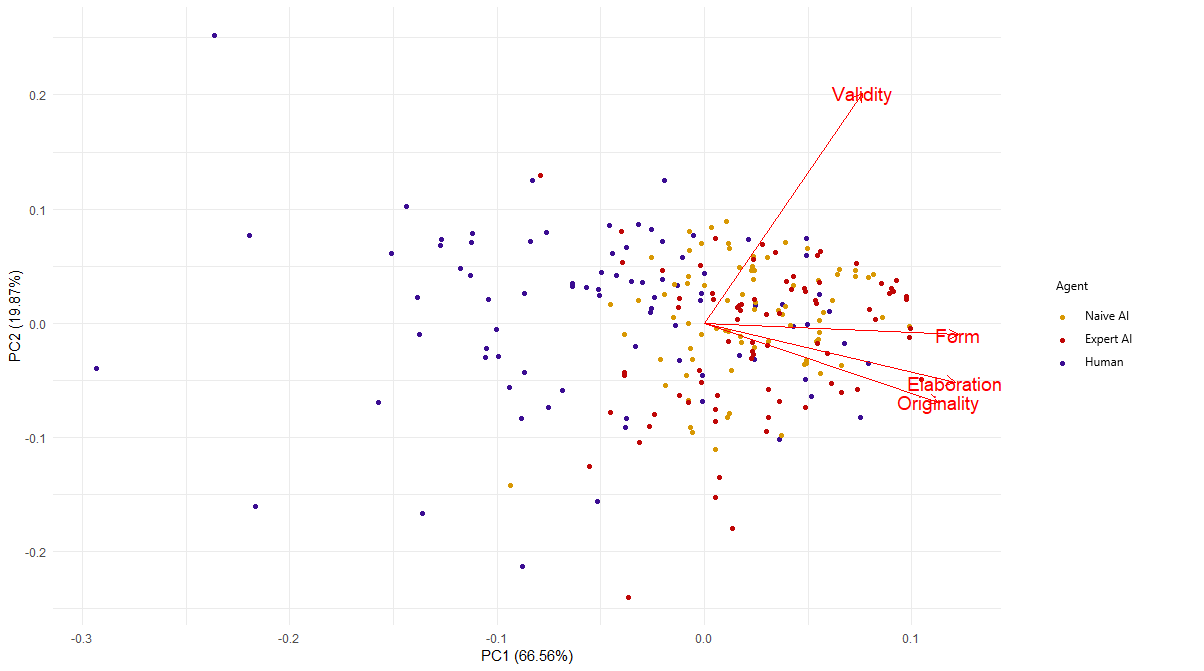}
    \label{fig:text_pca}
\end{figure}

\begin{table}[!htbp] \centering 
\renewcommand{\arraystretch}{1}
\setlength{\tabcolsep}{0.2pt} 
  \caption{Text Task PCA -- Human Evalutation}
  \label{} 
\scalebox{0.84}{
\begin{threeparttable}
\begin{tabular}{@{\extracolsep{5pt}}lrlr} 
\toprule
\midrule
name & Contribution dim 1 & name & Contribution dim 2\\
\midrule
Form & 31.15 & Validity & 84.28\\
Elaboration & 30.40 & Originality & 9.92\\
Originality & 26.46 & Elaboration & 5.59\\
Validity & 11.99 & Form & 0.21\\
\hline 
\hline \\[-1.8ex] 
\end{tabular} 
 \end{threeparttable}
 }
\end{table} 

\begin{figure}[h!]
    \centering
    \caption{}
    \includegraphics[width=0.75\linewidth]{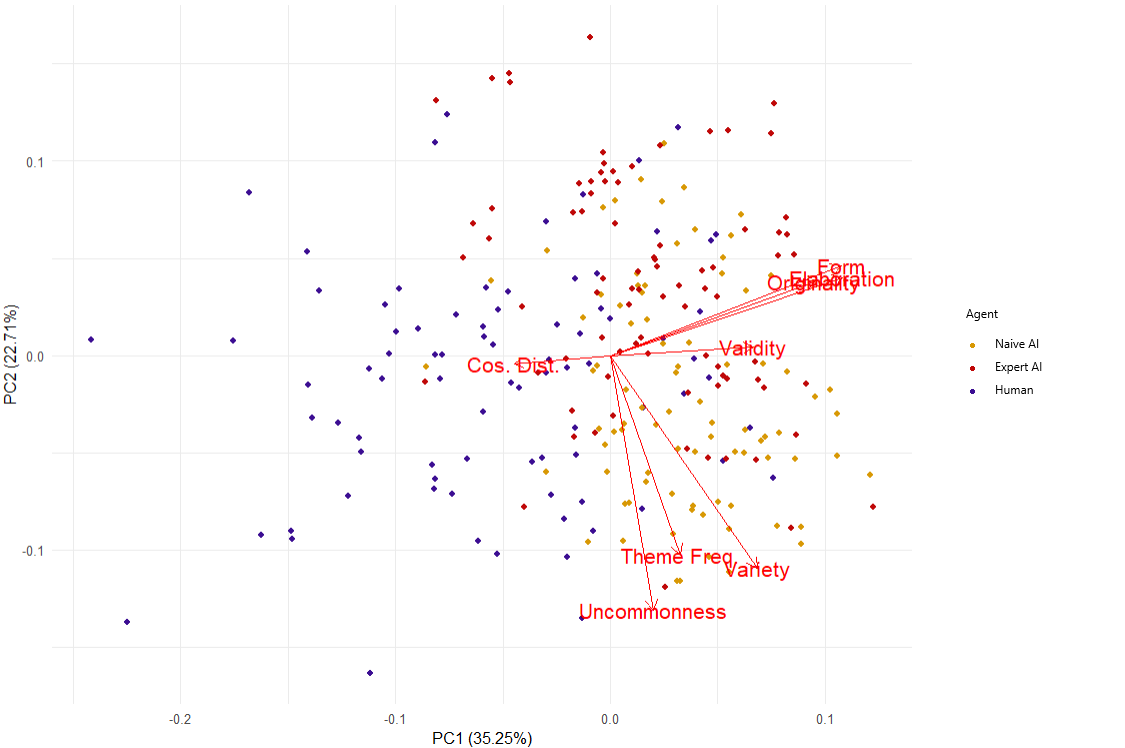}
    \label{fig:text_pca_all}
\end{figure}

\begin{table}[!htbp] \centering 
\renewcommand{\arraystretch}{1}
\setlength{\tabcolsep}{0.2pt} 
  \caption{Text Task PCA -- All Creativity Criteria}
  \label{} 
\scalebox{0.84}{
\begin{threeparttable}
\begin{tabular}{@{\extracolsep{5pt}}lrlr} 
\\[-1.8ex]\hline 
\hline \\[-1.8ex] 
name & Contribution dim 1 & name & Contribution dim 2\\
\midrule
Elaboration & 26.09 & Uncommonness & 38.14\\
Form & 25.68 & Variety & 26.74\\
Originality & 20.16 & Theme Freq. & 23.43\\
Variety & 10.42 & Form & 4.76\\
Validity & 9.93 & Elaboration & 3.60\\
\addlinespace
Cos. Dist. & 4.49 & Originality & 3.25\\
Theme Freq. & 2.35 & Validity & 0.04\\
Uncommonness & 0.88 & Cos. Dist. & 0.04\\
\hline 
\hline \\[-1.8ex] 
\end{tabular} 
 \end{threeparttable}
 }
\end{table} 

\subsubsection{Alternative Uses task}

\begin{figure}[H]
    \centering
    \caption{}
    \includegraphics[width=0.75\linewidth]{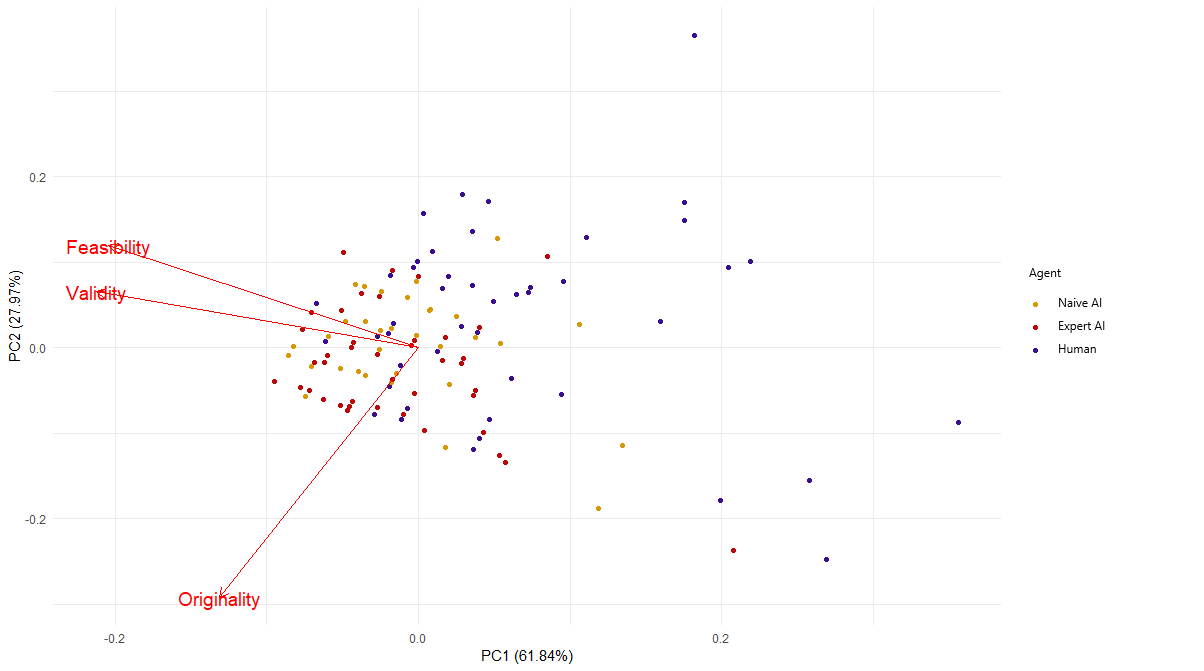}
    \label{fig:aut_pca}
\end{figure}

\begin{table}[H] \centering 
\renewcommand{\arraystretch}{1}
\setlength{\tabcolsep}{0.2pt} 
  \caption{Alternative Uses Task PCA -- Human Evalutation}
  \label{} 
\scalebox{0.84}{
\begin{threeparttable}
\begin{tabular}{@{\extracolsep{5pt}}lrlr} 
\toprule
\midrule
name & Contribution dim 1 & name & Contribution dim 2\\
\midrule
Validity & 43.40 & Originality & 82.11\\
Feasibility & 40.07 & Feasibility & 13.78\\
Originality & 16.54 & Validity & 4.11\\

\hline 
\hline \\[-1.8ex] 
\end{tabular} 
 \end{threeparttable}
 }
\end{table} 

\begin{figure}[H]
    \centering
    \caption{}
    \includegraphics[width=0.75\linewidth]{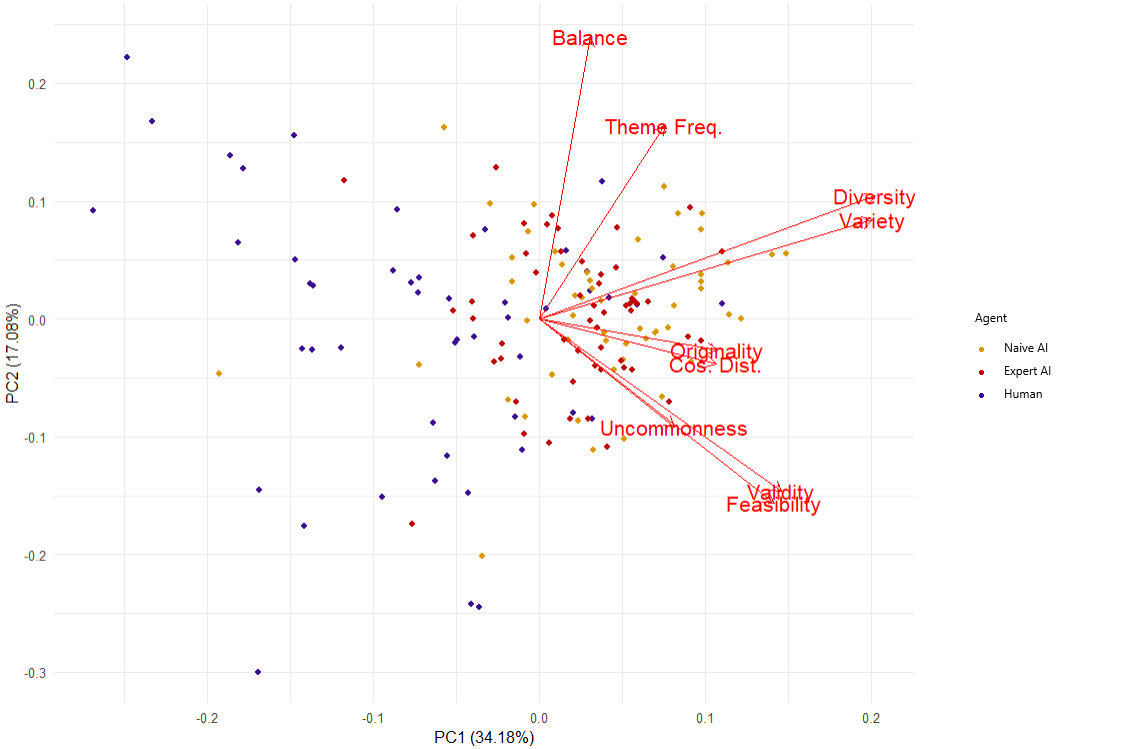}
    \label{fig:aut_pca_all}
\end{figure}

\begin{table}[!htbp] \centering 
\renewcommand{\arraystretch}{1}
\setlength{\tabcolsep}{0.2pt} 
  \caption{Alternative Uses Task PCA -- All Creativity Criteria}
  \label{} 
\scalebox{0.84}{
\begin{threeparttable}
\begin{tabular}{@{\extracolsep{5pt}}lrlr} 
\toprule
\midrule
name & Contribution dim 1 & name & Contribution dim 2\\
\midrule
Diversity & 25.83 & Balance & 36.14\\
Variety & 25.36 & Theme Freq. & 17.03\\
Validity & 13.44 & Feasibility & 15.41\\
Feasibility & 12.60 & Validity & 13.44\\
Originality & 7.24 & Diversity & 6.92\\
\addlinespace
Cos. Dist. & 7.18 & Uncommonness & 5.27\\
Uncommonness & 4.19 & Variety & 4.46\\
Theme Freq. & 3.56 & Cos. Dist. & 0.91\\
Balance & 0.59 & Originality & 0.43\\
\hline 
\hline \\[-1.8ex] 
\end{tabular} 
 \end{threeparttable}
 }
\end{table}

\subsubsection{Draw task}

\begin{figure}[H]
    \centering
    \caption{}
    \includegraphics[width=0.75\linewidth]{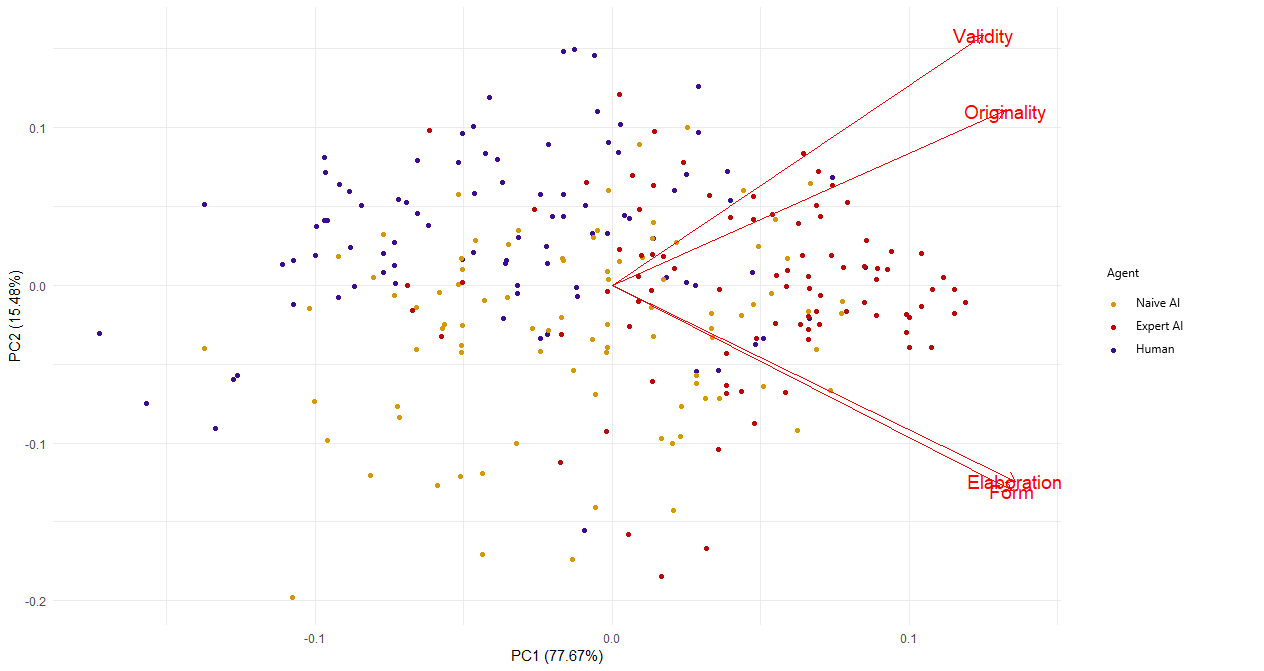}
    \label{fig:draw_pca}
\end{figure}

\begin{table}[!htbp] \centering 
\renewcommand{\arraystretch}{1}
\setlength{\tabcolsep}{0.2pt} 
  \caption{Drawing Task PCA -- Human Evalutation}
  \label{} 
\scalebox{0.84}{
\begin{threeparttable}
\begin{tabular}{@{\extracolsep{5pt}}lrlr} 
\toprule
\midrule
name & Contribution dim 1 & name & Contribution dim 2\\
\midrule
Elaboration & 26.50 & Validity & 36.01\\
Form & 26.02 & Form & 24.32\\
Originality & 25.09 & Elaboration & 22.13\\
Validity & 22.39 & Originality & 17.54\\

\hline 
\hline \\[-1.8ex] 
\end{tabular} 
 \end{threeparttable}
 }
\end{table} 

\begin{figure}[H]
    \centering
    \caption{}
    \includegraphics[width=0.75\linewidth]{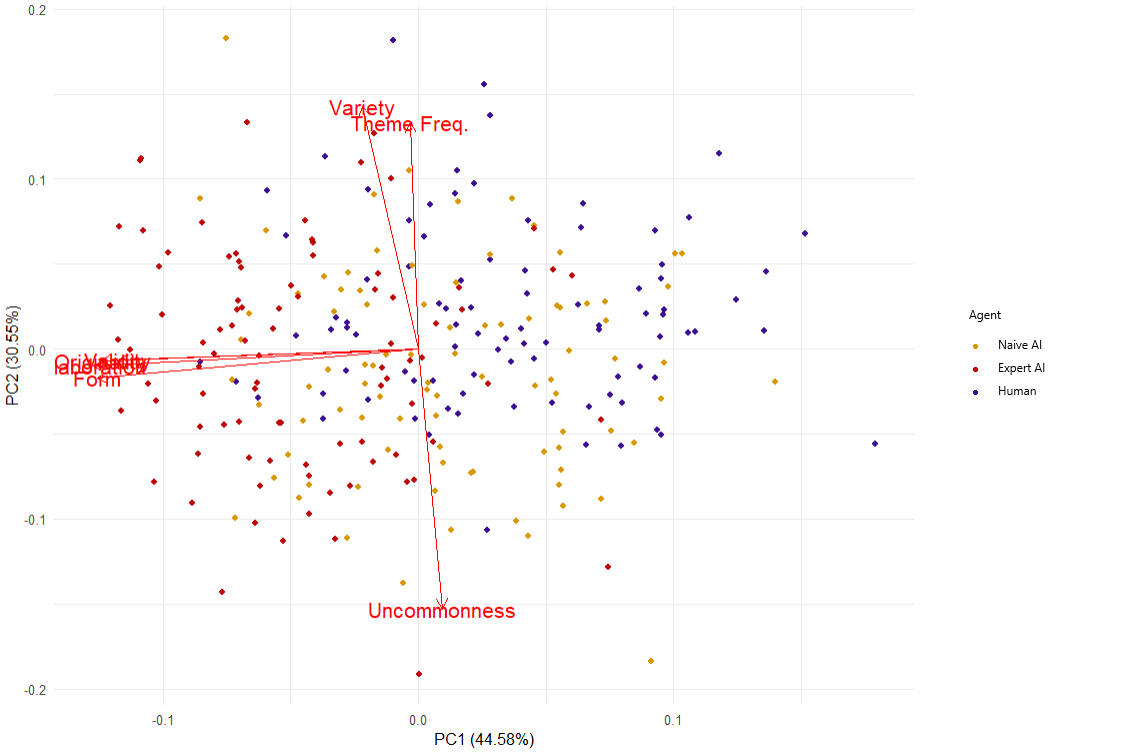}
    \label{fig:draw_pca_all}
\end{figure}

\begin{table}[!htbp] \centering 
\renewcommand{\arraystretch}{1}
\setlength{\tabcolsep}{0.2pt} 
  \caption{Drawing Task PCA -- Human Evalutation}
  \label{} 
\scalebox{0.84}{
\begin{threeparttable}
\begin{tabular}{@{\extracolsep{5pt}}lrlr} 
\toprule
\midrule
name & Contribution dim 1 & name & Contribution dim 2\\
\midrule
Elaboration & 26.21 & Uncommonness & 37.79\\
Form & 25.57 & Variety & 32.85\\
Originality & 24.93 & Theme Freq. & 28.57\\
Validity & 22.34 & Form & 0.47\\
Variety & 0.79 & Elaboration & 0.16\\
\addlinespace
Uncommonness & 0.15 & Originality & 0.07\\
Theme Freq. & 0.02 & Validity & 0.07\\
\hline 
\hline \\[-1.8ex] 
\end{tabular} 
 \end{threeparttable}
 }
\end{table} 
\clearpage

\subsection{Additional tests and regressions}

\subsubsection{Differences in Evaluators' Characteristics across tasks}

\begin{table}[!htbp] \centering 
\renewcommand{\arraystretch}{1}
\setlength{\tabcolsep}{0.2pt} 
  \caption{Socio-Demographic Variables Descriptive Statistics}
  \label{} 
\scalebox{1}{
\begin{threeparttable}
\begin{tabular}{@{\extracolsep{5pt}}lccc} 
\\[-1.8ex]\hline 
\hline \\[-1.8ex] 
Variable & Text & Alternative Uses & Draw\\
\midrule
\addlinespace
Mean Age & 22.814 (3.036) & 22.094 (2.646) & 22.031 (2.952)\\
Mean Gender & 0.657 (0.475) & 0.672 (0.47) & 0.831 (0.375)\\
Mean Diploma Licence & 0.414 (0.493) & 0.484 (0.5) & 0.415 (0.493)\\
Mean Diploma Master & 0.443 (0.497) & 0.469 (0.499) & 0.508 (0.5)\\
Mean Diploma Doctorat & 0.057 (0.232) & 0 (0) & 0.031 (0.173)\\
\addlinespace
Mean Droit & 0.057 (0.232) & 0.047 (0.212) & 0.015 (0.123)\\
Mean Economie Gestion & 0.414 (0.493) & 0.297 (0.457) & 0.4 (0.49)\\
Mean Lettres Langues & 0.057 (0.232) & 0.047 (0.212) & 0.046 (0.21)\\
Mean Sciences exactes & 0.186 (0.389) & 0.203 (0.403) & 0.154 (0.361)\\
Mean Psycho Socio & 0.071 (0.258) & 0.062 (0.242) & 0.092 (0.29)\\
\addlinespace
Mean Sciences politiques & 0.029 (0.167) & 0.094 (0.292) & 0.092 (0.29)\\
Mean Native speaker & 0.8 (0.4) & 0.906 (0.292) & 0.923 (0.267)\\
Mean Writting & 0.229 (0.42) & 0.219 (0.414) & 0.292 (0.455)\\
Mean Drawing & 0.286 (0.452) & 0.203 (0.403) & 0.2 (0.4)\\
Mean Painting & 0.157 (0.364) & 0.219 (0.414) & 0.138 (0.346)\\
\addlinespace
Mean Reading & 0.743 (0.437) & 0.766 (0.424) & 0.785 (0.411)\\
Mean Art & 0.1 (0.3) & 0.156 (0.363) & 0.2 (0.4)\\
Mean Cinema & 0.486 (0.5) & 0.562 (0.497) & 0.523 (0.5)\\
Mean French skills & 7.671 (1.481) & 8.25 (1.415) & 8.462 (1.039)\\
Mean Self Awareness & 0.414 (0.493) & 0.281 (0.45) & 0.308 (0.462)\\
\addlinespace
Mean Dectect AI & 0.114 (0.318) & 0.031 (0.174) & 0.092 (0.29)\\
Mean Jug favorable human & 0.029 (0.167) & 0 (0) & 0.031 (0.173)\\
Mean Jug favorable ai & 0.014 (0.119) & 0 (0) & 0 (0)\\
Mean Jug no impact & 0.071 (0.258) & 0.031 (0.174) & 0.062 (0.24)\\
Mean AI knowledge & 3.043 (0.801) & 2.762 (0.771) & 2.938 (0.927)\\
\addlinespace
Mean AI usage daily & 1.057 (0.827) & 1.175 (0.847) & 1.369 (0.97)\\
Mean AI app daily text & 0.882 (0.322) & 0.875 (0.331) & 0.87 (0.336)\\
Mean AI app daily image & 0.059 (0.235) & 0.083 (0.277) & 0.093 (0.29)\\
Mean AI app daily trad & 0.608 (0.489) & 0.562 (0.497) & 0.593 (0.492)\\
Mean AI usage work & 1.443 (0.822) & 1.476 (0.871) & 1.892 (0.914)\\
\addlinespace
Mean AI app work text & 0.933 (0.25) & 0.926 (0.262) & 0.934 (0.248)\\
Mean AI app work image & 0.017 (0.128) & 0.056 (0.229) & 0.016 (0.127)\\
Mean AI app work trad & 0.583 (0.493) & 0.574 (0.495) & 0.59 (0.492)\\
Mean Amelioration & 3.314 (0.919) & 3.254 (1.07) & 3.277 (1.075)\\
Mean Pertinence & 2.886 (0.767) & 3.159 (0.947) & 3.015 (0.985)\\
\addlinespace
Mean Self Creativity & 3.329 (0.751) & 3.286 (0.951) & 3.231 (0.941)\\
Mean AI Creativity & 0.514 (0.5) & 0.656 (0.475) & 0.538 (0.499)\\
Mean Willingness & 0.486 (0.5) & 0.375 (0.485) & 0.292 (0.455)\\
\addlinespace
\hline
\addlinespace
\# Evaluators & 70 & 64 & 65\\
\addlinespace
\hline 
\hline \\[-1.8ex] 
\end{tabular} 
 \end{threeparttable}
 }
\end{table} 
\clearpage
\begin{table}[!htbp] 
\centering 
\renewcommand{\arraystretch}{1}
\setlength{\tabcolsep}{0.2pt} 
  \caption{Comparison of Evaluators' Socio-Demographic Characteristics Across Tasks}
  \label{} 
\scalebox{1}{
\begin{threeparttable}
\begin{tabular}{@{\extracolsep{5pt}}lccc} 
\\[-1.8ex]\hline 
\hline \\[-1.8ex] 
Metric & Draw vs Alternative Uses & Text vs Alternative Uses & Text vs Draw\\
\midrule
\addlinespace
Age & -0.063 & \textcolor{blue}{0.721}$^{***}$ & \textcolor{blue}{0.784}$^{***}$\\
Gender & \textcolor{blue}{0.159}$^{***}$ & -0.015 & \textcolor{red}{-0.174}$^{***}$\\
Diploma Licence & \textcolor{red}{-0.069}$^{**}$ & \textcolor{red}{-0.07}$^{**}$ & -0.001\\
Diploma Master & 0.039 & -0.026 & \textcolor{red}{-0.065}$^{**}$\\
Diploma Doctorat & \textcolor{blue}{0.031}$^{***}$ & \textcolor{blue}{0.057}$^{***}$ & \textcolor{blue}{0.026}$^{***}$\\
\addlinespace
Droit & \textcolor{red}{-0.031}$^{***}$ & 0.01 & \textcolor{blue}{0.042}$^{***}$\\
Economie Gestion & \textcolor{blue}{0.103}$^{***}$ & \textcolor{blue}{0.117}$^{***}$ & 0.014\\
Lettres Langues & -0.001 & 0.01 & 0.011\\
Sciences exactes & \textcolor{red}{-0.049}$^{**}$ & -0.017 & 0.032\\
Psycho Socio & \textcolor{blue}{0.03}$^{*}$ & 0.009 & -0.021\\
\addlinespace
Sciences politiques & -0.001 & \textcolor{red}{-0.065}$^{***}$ & \textcolor{red}{-0.064}$^{***}$\\
Native speaker & 0.017 & \textcolor{red}{-0.106}$^{***}$ & \textcolor{red}{-0.123}$^{***}$\\
Writting & \textcolor{blue}{0.074}$^{***}$ & 0.01 & \textcolor{red}{-0.064}$^{***}$\\
Drawing & -0.003 & \textcolor{blue}{0.083}$^{***}$ & \textcolor{blue}{0.086}$^{***}$\\
Painting & \textcolor{red}{-0.08}$^{***}$ & \textcolor{red}{-0.062}$^{***}$ & 0.019\\
\addlinespace
Reading & 0.019 & -0.023 & \textcolor{red}{-0.042}$^{*}$\\
Art & \textcolor{blue}{0.044}$^{*}$ & \textcolor{red}{-0.056}$^{***}$ & \textcolor{red}{-0.1}$^{***}$\\
Cinema & -0.039 & \textcolor{red}{-0.077}$^{***}$ & -0.037\\
French skills & \textcolor{blue}{0.212}$^{***}$ & \textcolor{red}{-0.579}$^{***}$ & \textcolor{red}{-0.79}$^{***}$\\
Self Awareness & 0.026 & \textcolor{blue}{0.133}$^{***}$ & \textcolor{blue}{0.107}$^{***}$\\
\addlinespace
Detect AI & \textcolor{blue}{0.061}$^{***}$ & \textcolor{blue}{0.083}$^{***}$ & 0.022\\
Jug favorable human & \textcolor{blue}{0.031}$^{***}$ & \textcolor{blue}{0.029}$^{***}$ & -0.002\\
Jug favorable ai & 0 & \textcolor{blue}{0.014}$^{***}$ & \textcolor{blue}{0.014}$^{***}$\\
Jug no impact & \textcolor{blue}{0.03}$^{**}$ & \textcolor{blue}{0.04}$^{***}$ & 0.01\\
AI knowledge & \textcolor{blue}{0.177}$^{***}$ & \textcolor{blue}{0.281}$^{***}$ & \textcolor{blue}{0.104}$^{**}$\\
\addlinespace
AI usage daily & \textcolor{blue}{0.195}$^{***}$ & \textcolor{red}{-0.117}$^{**}$ & \textcolor{red}{-0.312}$^{***}$\\
AI app daily text & -0.005 & 0.007 & 0.012\\
AI app daily image & 0.009 & -0.025 & \textcolor{red}{-0.034}$^{**}$\\
AI app daily trad & 0.03 & 0.045 & 0.015\\
AI usage work & \textcolor{blue}{0.416}$^{***}$ & -0.033 & \textcolor{red}{-0.449}$^{***}$\\
\addlinespace
AI app work text & 0.009 & 0.007 & -0.001\\
AI app work image & \textcolor{red}{-0.039}$^{***}$ & \textcolor{red}{-0.039}$^{***}$ & 0\\
AI app work trad & 0.016 & 0.009 & -0.007\\
Amelioration & 0.023 & 0.06 & 0.037\\
Pertinence & \textcolor{red}{-0.143}$^{***}$ & \textcolor{red}{-0.273}$^{***}$ & \textcolor{red}{-0.13}$^{***}$\\
\addlinespace
Self Creativity & -0.055 & 0.043 & \textcolor{blue}{0.098}$^{**}$\\
AI Creativity & \textcolor{red}{-0.118}$^{***}$ & \textcolor{red}{-0.142}$^{***}$ & -0.024\\
Willingness & \textcolor{red}{-0.083}$^{***}$ & \textcolor{blue}{0.111}$^{***}$ & \textcolor{blue}{0.193}$^{***}$\\
\addlinespace
\hline 
\hline \\[-1.8ex] 
\end{tabular} 
\begin{tablenotes}
\footnotesize
\item \justifying {\it Notes:} This table presents the results of the ANOVA and Tukey's HSD test, and Pairwise Wilcoxon test for the specified variable and group. The coefficients represent the Mean differences between groups. ***, **, and * indicate significance at the 1\%, 5\%, and 10\% levels, respectively.
\end{tablenotes}
 \end{threeparttable}
 }
\end{table} 

\clearpage
\begin{table}[!htbp] 
\centering 
\renewcommand{\arraystretch}{1}
\setlength{\tabcolsep}{0.2pt} 
  \caption{Comparison of Evaluators' Socio-Demographic Characteristics Across Tasks}
  \label{} 
\scalebox{1}{
\begin{threeparttable}
\begin{tabular}{@{\extracolsep{5pt}}lccc} 
\\[-1.8ex]\hline 
\hline \\[-1.8ex] 
Metric & Draw vs Alternative Uses & Text vs Alternative Uses & Text vs Draw\\
\midrule
\addlinespace
Age & -0.063 & \textcolor{blue}{0.721}$^{***}$ & \textcolor{blue}{0.784}$^{***}$\\
Gender & \textcolor{blue}{0.159}$^{***}$ & -0.015 & \textcolor{red}{-0.174}$^{***}$\\
Diploma Licence & \textcolor{red}{-0.069}$^{**}$ & \textcolor{red}{-0.07}$^{**}$ & -0.001\\
Diploma Master & 0.039 & -0.026 & \textcolor{red}{-0.065}$^{**}$\\
Diploma Doctorat & \textcolor{blue}{0.031}$^{***}$ & \textcolor{blue}{0.057}$^{***}$ & \textcolor{blue}{0.026}$^{***}$\\
\addlinespace
Droit & \textcolor{red}{-0.031}$^{***}$ & 0.01 & \textcolor{blue}{0.042}$^{***}$\\
Economie Gestion & \textcolor{blue}{0.103}$^{***}$ & \textcolor{blue}{0.117}$^{***}$ & 0.014\\
Lettres Langues & -0.001 & 0.01 & 0.011\\
Sciences exactes & \textcolor{red}{-0.049}$^{**}$ & -0.017 & 0.032\\
Psycho Socio & \textcolor{blue}{0.03}$^{*}$ & 0.009 & -0.021\\
\addlinespace
Sciences politiques & -0.001 & \textcolor{red}{-0.065}$^{***}$ & \textcolor{red}{-0.064}$^{***}$\\
Native speaker & 0.017 & \textcolor{red}{-0.106}$^{***}$ & \textcolor{red}{-0.123}$^{***}$\\
Writting & \textcolor{blue}{0.074}$^{***}$ & 0.01 & \textcolor{red}{-0.064}$^{***}$\\
Drawing & -0.003 & \textcolor{blue}{0.083}$^{***}$ & \textcolor{blue}{0.086}$^{***}$\\
Painting & \textcolor{red}{-0.08}$^{***}$ & \textcolor{red}{-0.062}$^{***}$ & 0.019\\
\addlinespace
Reading & 0.019 & -0.023 & \textcolor{red}{-0.042}$^{*}$\\
Art & \textcolor{blue}{0.044}$^{*}$ & \textcolor{red}{-0.056}$^{***}$ & \textcolor{red}{-0.1}$^{***}$\\
Cinema & -0.039 & \textcolor{red}{-0.077}$^{**}$ & -0.037\\
French skills & \textcolor{blue}{0.212}$^{***}$ & \textcolor{red}{-0.579}$^{***}$ & \textcolor{red}{-0.79}$^{***}$\\
Self Awareness & 0.026 & \textcolor{blue}{0.133}$^{***}$ & \textcolor{blue}{0.107}$^{***}$\\
\addlinespace
Detect AI & \textcolor{blue}{0.061}$^{***}$ & \textcolor{blue}{0.083}$^{***}$ & 0.022\\
Jug favorable human & \textcolor{blue}{0.031}$^{***}$ & \textcolor{blue}{0.029}$^{***}$ & -0.002\\
Jug favorable ai & 0 & \textcolor{blue}{0.014}$^{***}$ & \textcolor{blue}{0.014}$^{***}$\\
Jug no impact & \textcolor{blue}{0.03}$^{**}$ & \textcolor{blue}{0.04}$^{***}$ & 0.01\\
AI knowledge & \textcolor{blue}{0.177}$^{***}$ & \textcolor{blue}{0.281}$^{***}$ & \textcolor{blue}{0.104}$^{**}$\\
\addlinespace
AI usage daily & \textcolor{blue}{0.195}$^{***}$ & \textcolor{red}{-0.117}$^{**}$ & \textcolor{red}{-0.312}$^{***}$\\
AI app daily text & -0.005 & 0.007 & 0.012\\
AI app daily image & 0.009 & -0.025 & \textcolor{red}{-0.034}$^{**}$\\
AI app daily trad & 0.03 & 0.045 & 0.015\\
AI usage work & \textcolor{blue}{0.416}$^{***}$ & -0.033 & \textcolor{red}{-0.449}$^{***}$\\
\addlinespace
AI app work text & 0.009 & 0.007 & -0.001\\
AI app work image & \textcolor{red}{-0.039}$^{***}$ & \textcolor{red}{-0.039}$^{***}$ & 0\\
AI app work trad & 0.016 & 0.009 & -0.007\\
Amelioration & 0.023 & 0.06 & 0.037\\
Pertinence & \textcolor{red}{-0.143}$^{***}$ & \textcolor{red}{-0.273}$^{***}$ & \textcolor{red}{-0.13}$^{***}$\\
\addlinespace
Self Creativity & -0.055 & 0.043 & \textcolor{blue}{0.098}$^{**}$\\
AI Creativity & \textcolor{red}{-0.118}$^{***}$ & \textcolor{red}{-0.142}$^{***}$ & -0.024\\
Willingness & \textcolor{red}{-0.083}$^{***}$ & \textcolor{blue}{0.111}$^{***}$ & \textcolor{blue}{0.193}$^{***}$\\
\addlinespace
\hline 
\hline \\[-1.8ex] 
\end{tabular} 
\begin{tablenotes}
\footnotesize
\item \justifying {\it Notes:} This table presents the results of the ANOVA and Tukey's HSD test test for the specified variable and group. The coefficients represent the Mean differences between groups. ***, **, and * indicate significance at the 1\%, 5\%, and 10\% levels, respectively.
\end{tablenotes}
 \end{threeparttable}
 }
\end{table} 

\clearpage

\subsubsection{Effect of Evaluators' Characteristics on Evaluation}

\subsubsection{Text task}

\begin{table}[!htbp] \centering 
\renewcommand{\arraystretch}{0.8}
\setlength{\tabcolsep}{0.05pt} 
  \caption{OLS -- Average Score by Evaluators} 
  \label{} 
\scalebox{1}{
\begin{threeparttable}
\begin{tabular}{@{\extracolsep{5pt}}lcccc} 
\\[-1.8ex]\hline 
\hline \\[-1.8ex] 
 & \multicolumn{4}{c}{\textit{Dependent variable:}} \\ 
\cline{2-5} 
\\[-1.8ex] & Validity & Form & Elaboration & Originality \\ 
\\[-1.8ex] & (1) & (2) & (3) & (4)\\ 
\hline \\[-1.8ex] 
\addlinespace
 Age & 0.024 & $-$0.023 & $-$0.024 & 0.011 \\ 
  & (0.016) & (0.019) & (0.021) & (0.020) \\ 
  & & & & \\ 
 Gender & 0.042 & 0.233$^{*}$ & 0.139 & 0.298$^{**}$ \\ 
  & (0.100) & (0.122) & (0.132) & (0.127) \\ 
  & & & & \\ 
 Native speaker & $-$0.183 & $-$0.503$^{***}$ & $-$0.296$^{*}$ & $-$0.329$^{**}$ \\ 
  & (0.120) & (0.147) & (0.159) & (0.153) \\ 
  & & & & \\ 
 Detect AI & $-$0.389$^{**}$ & 0.091 & 0.180 & 0.067 \\ 
  & (0.175) & (0.214) & (0.232) & (0.223) \\ 
  & & & & \\ 
 Jug favorable human & 0.317 & $-$0.166 & $-$0.167 & $-$0.730$^{*}$ \\ 
  & (0.333) & (0.408) & (0.442) & (0.425) \\ 
  & & & & \\ 
 Jug favorable ai & 1.002$^{**}$ & 0.187 & $-$0.271 & $-$0.197 \\ 
  & (0.436) & (0.534) & (0.579) & (0.556) \\ 
  & & & & \\ 
 AI knowledge & $-$0.068 & $-$0.063 & $-$0.087 & $-$0.031 \\ 
  & (0.062) & (0.076) & (0.082) & (0.079) \\ 
  & & & & \\ 
 AI usage daily & $-$0.083 & $-$0.162$^{**}$ & $-$0.219$^{**}$ & $-$0.168$^{**}$ \\ 
  & (0.065) & (0.079) & (0.086) & (0.083) \\ 
  & & & & \\ 
 AI usage work & 0.243$^{***}$ & 0.146 & 0.081 & 0.052 \\ 
  & (0.079) & (0.097) & (0.106) & (0.101) \\ 
  & & & & \\ 
 Amelioration & $-$0.052 & 0.053 & 0.181$^{**}$ & 0.0001 \\ 
  & (0.069) & (0.084) & (0.091) & (0.088) \\ 
  & & & & \\ 
 Pertinence & 0.139$^{**}$ & 0.113 & 0.198$^{**}$ & 0.325$^{***}$ \\ 
  & (0.067) & (0.082) & (0.088) & (0.085) \\ 
  & & & & \\ 
 AI Creativity & $-$0.144 & 0.105 & 0.145 & 0.111 \\ 
  & (0.095) & (0.117) & (0.127) & (0.122) \\ 
  & & & & \\ 
 Willingness & $-$0.146 & $-$0.161 & $-$0.214 & $-$0.126 \\ 
  & (0.101) & (0.123) & (0.134) & (0.128) \\ 
  & & & & \\ 
 Constant & 3.722$^{***}$ & 4.163$^{***}$ & 3.565$^{***}$ & 2.509$^{***}$ \\ 
  & (0.434) & (0.531) & (0.577) & (0.553) \\ 
  & & & & \\ 
\hline \\[-1.8ex] 
Observations & 210 & 210 & 210 & 210 \\ 
R$^{2}$ & 0.125 & 0.145 & 0.146 & 0.171 \\ 
Adjusted R$^{2}$ & 0.067 & 0.088 & 0.089 & 0.116 \\ 
Residual Std. Error (df = 196) & 0.629 & 0.769 & 0.834 & 0.801 \\ 
F Statistic (df = 13; 196) & 2.162$^{**}$ & 2.559$^{***}$ & 2.568$^{***}$ & 3.119$^{***}$ \\ 
\hline 
\hline \\[-1.8ex] 
\end{tabular} 
\begin{tablenotes}
 \footnotesize
 \item \justifying   {\it Notes:}This table presents the coefficients reflecting the impact of evaluators' socio-demographic characteristics on the various creativity criteria scores. ***, ** and * indicate significance at the 1\%, 5\% and 10\% level, respectively. Effects are estimated using a OLS. \end{tablenotes}
 \end{threeparttable}
 }
\end{table} 
\begin{table}[!htbp] \centering 
\renewcommand{\arraystretch}{0.8}
\setlength{\tabcolsep}{0.05pt} 
  \caption{Polynomial Logit -- Evaluators Profiles} 
  \label{} 
\scalebox{1}{
\begin{threeparttable}
\begin{tabular}{@{\extracolsep{5pt}}lcccc} 
\\[-1.8ex]\hline 
\hline \\[-1.8ex] 
 & \multicolumn{4}{c}{\textit{Dependent variable:}} \\ 
\cline{2-5} 
\\[-1.8ex] & Validity & Form & Elaboration & Originality \\ 
\\[-1.8ex] & (1) & (2) & (3) & (4)\\ 
\hline \\[-1.8ex] 
\addlinespace
 \# Words & 0.002$^{***}$ & 0.005$^{***}$ & 0.010$^{***}$ & 0.005$^{***}$ \\ 
  & (0.001) & (0.001) & (0.001) & (0.001) \\ 
  & & & & \\ 
 Age & 0.039$^{*}$ & $-$0.049$^{**}$ & $-$0.063$^{***}$ & 0.011 \\ 
  & (0.022) & (0.022) & (0.021) & (0.020) \\ 
  & & & & \\ 
 Gender & $-$0.197 & 0.344$^{**}$ & 0.191 & 0.469$^{***}$ \\ 
  & (0.145) & (0.136) & (0.139) & (0.135) \\ 
  & & & & \\ 
 Native speaker & $-$0.220 & $-$0.811$^{***}$ & $-$0.505$^{***}$ & $-$0.441$^{***}$ \\ 
  & (0.166) & (0.163) & (0.167) & (0.162) \\ 
  & & & & \\ 
 Detect AI & $-$0.816$^{***}$ & 0.320 & 0.535$^{**}$ & 0.281 \\ 
  & (0.239) & (0.254) & (0.258) & (0.244) \\ 
  & & & & \\ 
 Jug favorable human & 0.526 & $-$0.574 & $-$0.612 & $-$1.436$^{***}$ \\ 
  & (0.465) & (0.464) & (0.445) & (0.430) \\ 
  & & & & \\ 
 Jug favorable ai & 3.552$^{***}$ & 0.283 & $-$0.918 & $-$0.602 \\ 
  & (1.092) & (0.604) & (0.559) & (0.533) \\ 
  & & & & \\ 
 AI knowledge & $-$0.181$^{**}$ & $-$0.128 & $-$0.200$^{**}$ & $-$0.102 \\ 
  & (0.088) & (0.083) & (0.085) & (0.083) \\ 
  & & & & \\ 
 AI usage daily & $-$0.150 & $-$0.326$^{***}$ & $-$0.471$^{***}$ & $-$0.232$^{***}$ \\ 
  & (0.092) & (0.089) & (0.090) & (0.086) \\ 
  & & & & \\ 
 AI usage work & 0.553$^{***}$ & 0.356$^{***}$ & 0.279$^{**}$ & 0.143 \\ 
  & (0.113) & (0.109) & (0.109) & (0.105) \\ 
  & & & & \\ 
 Amelioration & $-$0.157 & 0.146 & 0.400$^{***}$ & 0.076 \\ 
  & (0.098) & (0.093) & (0.094) & (0.091) \\ 
  & & & & \\ 
 Pertinence & 0.333$^{***}$ & 0.232$^{**}$ & 0.454$^{***}$ & 0.496$^{***}$ \\ 
  & (0.096) & (0.093) & (0.093) & (0.090) \\ 
  & & & & \\ 
 AI Creativity & $-$0.371$^{***}$ & 0.125 & 0.206 & 0.096 \\ 
  & (0.138) & (0.129) & (0.132) & (0.128) \\ 
  & & & & \\ 
 Willingness & $-$0.238 & $-$0.481$^{***}$ & $-$0.650$^{***}$ & $-$0.346$^{**}$ \\ 
  & (0.145) & (0.142) & (0.144) & (0.139) \\ 
  & & & & \\ 
\hline \\[-1.8ex] 
AIC & 2221 & 2659.55 & 2511.37 & 2916.83 \\ 
Observations & 980 & 980 & 980 & 980 \\ 
\hline 
\hline \\[-1.8ex] 
\end{tabular} 
\begin{tablenotes}
 \footnotesize
 \item \justifying   {\it Notes:}This table presents the coefficients reflecting the impact of evaluators' socio-demographic characteristics on the various creativity criteria scores. ***, ** and * indicate significance at the 1\%, 5\% and 10\% level, respectively. Effects are estimated using an Ordered Polynomial Logit. \end{tablenotes}
 \end{threeparttable}
 }
\end{table} 
\clearpage

\subsubsection{Alternative Uses task}

\begin{table}[!htbp] \centering 
\renewcommand{\arraystretch}{0.8}
\setlength{\tabcolsep}{0.05pt} 
  \caption{OLS -- Average Score by Evaluators} 
  \label{} 
\scalebox{1}{
\begin{threeparttable}
\begin{tabular}{@{\extracolsep{5pt}}lccc}  
\\[-1.8ex]\hline 
\hline \\[-1.8ex] 
 & \multicolumn{3}{c}{\textit{Dependent variable:}} \\ 
\cline{2-4} 
\\[-1.8ex] & Validity Max & Feasibility Max & Originality Max \\
\hline \\[-1.8ex] 
\addlinespace
 Age & $-$0.061$^{**}$ & $-$0.036 & 0.002 \\ 
  & (0.024) & (0.022) & (0.016) \\ 
  & & & \\ 
 Gender & $-$0.169 & $-$0.232$^{*}$ & $-$0.012 \\ 
  & (0.133) & (0.120) & (0.090) \\ 
  & & & \\ 
 Native speaker & $-$0.108 & $-$0.494$^{**}$ & $-$0.125 \\ 
  & (0.218) & (0.198) & (0.148) \\ 
  & & & \\ 
 Detect AI & 0.469 & 0.688$^{**}$ & 0.350 \\ 
  & (0.374) & (0.339) & (0.253) \\ 
  & & & \\ 
 AI knowledge & 0.109 & 0.063 & $-$0.010 \\ 
  & (0.093) & (0.084) & (0.063) \\ 
  & & & \\ 
 AI usage daily & $-$0.102 & $-$0.304$^{***}$ & $-$0.090 \\ 
  & (0.106) & (0.096) & (0.072) \\ 
  & & & \\ 
 AI usage work & $-$0.027 & 0.086 & $-$0.062 \\ 
  & (0.104) & (0.094) & (0.070) \\ 
  & & & \\ 
 Amelioration & $-$0.115 & 0.016 & $-$0.013 \\ 
  & (0.085) & (0.077) & (0.057) \\ 
  & & & \\ 
 Pertinence & 0.271$^{***}$ & 0.003 & 0.087 \\ 
  & (0.083) & (0.076) & (0.056) \\ 
  & & & \\ 
 AI Creativity & $-$0.163 & $-$0.004 & $-$0.171$^{*}$ \\ 
  & (0.140) & (0.127) & (0.095) \\ 
  & & & \\ 
 Willingness & 0.132 & 0.265$^{*}$ & $-$0.045 \\ 
  & (0.151) & (0.137) & (0.102) \\ 
  & & & \\ 
 Constant & 5.330$^{***}$ & 5.564$^{***}$ & 4.766$^{***}$ \\ 
  & (0.670) & (0.608) & (0.453) \\ 
  & & & \\ 
\hline \\[-1.8ex] 
Observations & 189 & 189 & 189 \\ 
R$^{2}$ & 0.127 & 0.124 & 0.101 \\ 
Adjusted R$^{2}$ & 0.072 & 0.069 & 0.045 \\ 
Residual Std. Error (df = 177) & 0.773 & 0.701 & 0.522 \\ 
F Statistic (df = 11; 177) & 2.332$^{**}$ & 2.267$^{**}$ & 1.806$^{*}$ \\ 
\hline 
\hline \\[-1.8ex] 
\end{tabular} 
\begin{tablenotes}
 \footnotesize
 \item \justifying   {\it Notes:}This table presents the coefficients reflecting the impact of evaluators' socio-demographic characteristics on the various creativity criteria scores. ***, ** and * indicate significance at the 1\%, 5\% and 10\% level, respectively. Effects are estimated using a OLS. \end{tablenotes}
 \end{threeparttable}
 }
\end{table} 

\begin{table}[!htbp] \centering 
\renewcommand{\arraystretch}{0.8}
\setlength{\tabcolsep}{0.05pt} 
  \caption{Polynomial Logit -- Evaluators Profiles} 
  \label{} 
\scalebox{0.80}{
\begin{threeparttable}
\begin{tabular}{@{\extracolsep{5pt}}lccc} 
\\[-1.8ex]\hline 
\hline \\[-1.8ex] 
 & \multicolumn{3}{c}{\textit{Dependent variable:}} \\ 
\cline{2-4} 
\\[-1.8ex] & Validity Max & Feasibility Max & Originality Max \\ 
\\[-1.8ex] & (1) & (2) & (3)\\ 
\hline \\[-1.8ex] 
\addlinespace
 Mean \# Words & $-$0.016  & $-$0.006  & 0.022  \\ 
  & (0.036)  & (0.041) & (0.035)  \\ 
  & & & \\ 
 Age & $-$0.123$^{***}$  & 0.026 & $-$0.065$^{*}$  \\ 
  & (0.037)  & (0.047) & (0.035) \\ 
 & & & \\ 
 Gender & $-$0.306  & 0.070 & $-$0.397$^{**}$  \\ 
  & (0.199) & (0.223) & (0.196)  \\ 
 & & & \\ 
 Native speaker & $-$0.029  & 0.092 & $-$0.918$^{***}$  \\ 
  & (0.330)  & (0.381) & (0.343) \\ 
 & & & \\ 
 Detect AI & 2.054$^{*}$  & 1.465 & 0.785 \\ 
  & (1.081)  & (1.104) & (0.524) \\ 
 & & & \\ 
 AI knowledge & 0.144  & $-$0.018  & 0.089 \\ 
  & (0.137)  & (0.168)  & (0.136) \\ 
 & & & \\ 
 AI usage daily & $-$0.374$^{**}$ & $-$0.228  & $-$0.572$^{***}$ \\ 
  & (0.170)  & (0.187)  & (0.161) \\ 
 & & & \\ 
 AI usage work & 0.080  & $-$0.222 & 0.292$^{*}$ \\ 
  & (0.171)  & (0.185)  & (0.156) \\ 
 & & & \\ 
 Amelioration & $-$0.334$^{**}$ & $-$0.106  & $-$0.054 \\ 
  & (0.130)  & (0.152)  & (0.125) \\ 
 & & & \\ 
 Pertinence & 0.738$^{***}$  & 0.400$^{***}$  & 0.095 \\ 
  & (0.131)  & (0.153)  & (0.122) \\ 
 & & & \\ 
 AI Creativity & $-$0.434$^{*}$  & $-$0.658$^{**}$  & $-$0.168 \\ 
  & (0.222)  & (0.273) & (0.211) \\ 
 & & & \\ 
 Willingness & 0.261 & $-$0.092 & 0.529$^{**}$ \\ 
  & (0.225)  & (0.253)  & (0.222) \\ 
 & & & \\ 
\hline \\[-1.8ex] 
AIC & 1203.97 & 966.24 & 1298.24 \\ 
Observations & 567 & 567 & 567 \\ 
\hline 
\hline \\[-1.8ex] 
\end{tabular} 
\begin{tablenotes}
 \footnotesize
 \item \justifying   {\it Notes:}This table presents the coefficients reflecting the impact of evaluators' socio-demographic characteristics on the various creativity criteria scores. ***, ** and * indicate significance at the 1\%, 5\% and 10\% level, respectively. Effects are estimated using an Ordered Polynomial Logit. \end{tablenotes}
\end{threeparttable}
}
\end{table} 
\clearpage

\subsubsection{Draw task}

\begin{table}[!htbp] \centering 
\renewcommand{\arraystretch}{0.8}
\setlength{\tabcolsep}{0.05pt} 
  \caption{OLS -- Average Score by Evaluators} 
  \label{} 
\scalebox{1}{
\begin{threeparttable}
\begin{tabular}{@{\extracolsep{5pt}}lcccc} 
\\[-1.8ex]\hline 
\hline \\[-1.8ex] 
 & \multicolumn{4}{c}{\textit{Dependent variable:}} \\ 
\cline{2-5} 
\\[-1.8ex] & Validity & Form & Elaboration & Originality \\ 
\\[-1.8ex] & (1) & (2) & (3) & (4)\\ 
\hline \\[-1.8ex] 
\addlinespace
 Age & $-$0.092$^{***}$ & $-$0.044 & $-$0.062$^{**}$ & $-$0.043 \\ 
  & (0.026) & (0.028) & (0.029) & (0.027) \\ 
  & & & & \\ 
 Gender & 0.115 & 0.212 & 0.039 & 0.064 \\ 
  & (0.211) & (0.228) & (0.230) & (0.220) \\ 
  & & & & \\ 
 Native speaker & 0.701$^{***}$ & 0.863$^{***}$ & 0.675$^{**}$ & 0.720$^{***}$ \\ 
  & (0.261) & (0.281) & (0.284) & (0.272) \\ 
  & & & & \\ 
 Detect AI & $-$0.281 & $-$0.183 & $-$0.573$^{*}$ & $-$0.163 \\ 
  & (0.292) & (0.315) & (0.318) & (0.304) \\ 
  & & & & \\ 
 Jug favorable human & $-$0.828$^{*}$ & $-$1.183$^{**}$ & $-$0.403 & $-$0.909$^{*}$ \\ 
  & (0.490) & (0.529) & (0.533) & (0.511) \\ 
  & & & & \\ 
 AI knowledge & $-$0.063 & $-$0.009 & $-$0.070 & $-$0.0004 \\ 
  & (0.092) & (0.100) & (0.101) & (0.096) \\ 
  & & & & \\ 
 AI usage daily & 0.106 & 0.156$^{*}$ & 0.146 & 0.136 \\ 
  & (0.084) & (0.091) & (0.091) & (0.088) \\ 
  & & & & \\ 
 AI usage work & $-$0.128 & $-$0.061 & $-$0.086 & $-$0.179$^{*}$ \\ 
  & (0.094) & (0.101) & (0.102) & (0.098) \\ 
  & & & & \\ 
 Amelioration & 0.148 & 0.055 & 0.140 & 0.088 \\ 
  & (0.101) & (0.110) & (0.110) & (0.106) \\ 
  & & & & \\ 
 Pertinence & $-$0.059 & $-$0.020 & $-$0.001 & $-$0.052 \\ 
  & (0.081) & (0.088) & (0.089) & (0.085) \\ 
  & & & & \\ 
 AI Creativity & $-$0.384$^{**}$ & $-$0.179 & $-$0.160 & 0.006 \\ 
  & (0.174) & (0.188) & (0.189) & (0.181) \\ 
  & & & & \\ 
 Willingness & $-$0.002 & 0.118 & 0.008 & $-$0.187 \\ 
  & (0.195) & (0.210) & (0.212) & (0.203) \\ 
  & & & & \\ 
 Constant & 4.947$^{***}$ & 2.868$^{***}$ & 3.310$^{***}$ & 3.311$^{***}$ \\ 
  & (0.738) & (0.797) & (0.803) & (0.769) \\ 
  & & & & \\ 
\hline \\[-1.8ex] 
Observations & 195 & 195 & 195 & 195 \\ 
R$^{2}$ & 0.144 & 0.141 & 0.117 & 0.118 \\ 
Adjusted R$^{2}$ & 0.087 & 0.085 & 0.059 & 0.060 \\ 
Residual Std. Error (df = 182) & 0.917 & 0.991 & 0.999 & 0.956 \\ 
F Statistic (df = 12; 182) & 2.547$^{***}$ & 2.494$^{***}$ & 2.008$^{**}$ & 2.023$^{**}$ \\ 
\hline 
\hline \\[-1.8ex] 
\end{tabular} 
\begin{tablenotes}
 \footnotesize
 \item \justifying   {\it Notes:}This table presents the coefficients reflecting the impact of evaluators' socio-demographic characteristics on the various creativity criteria scores. ***, ** and * indicate significance at the 1\%, 5\% and 10\% level, respectively. Effects are estimated using a OLS. \end{tablenotes}
 \end{threeparttable}
 }
\end{table} 
\begin{table}[!htbp] \centering 
\renewcommand{\arraystretch}{0.8}
\setlength{\tabcolsep}{0.05pt} 
  \caption{Polynomial Logit -- Evaluators Profiles} 
  \label{} 
\scalebox{1}{
\begin{threeparttable}
\begin{tabular}{@{\extracolsep{5pt}}lcccc} 
\\[-1.8ex]\hline 
\hline \\[-1.8ex] 
 & \multicolumn{4}{c}{\textit{Dependent variable:}} \\ 
\cline{2-5} 
\\[-1.8ex] & Validity & Form & Elaboration & Originality \\ 
\\[-1.8ex] & (1) & (2) & (3) & (4)\\ 
\hline \\[-1.8ex] 
\addlinespace
 \% Pixel Cov. & 1.550$^{***}$ & 5.028$^{***}$ & 4.088$^{***}$ & 3.273$^{***}$ \\ 
  & (0.512) & (0.537) & (0.522) & (0.505) \\ 
  & & & & \\ 
 Age & $-$0.129$^{***}$ & $-$0.072$^{***}$ & $-$0.083$^{***}$ & $-$0.064$^{***}$ \\ 
  & (0.024) & (0.023) & (0.023) & (0.023) \\ 
  & & & & \\ 
 Gender & 0.146 & 0.226 & $-$0.016 & 0.021 \\ 
  & (0.178) & (0.179) & (0.180) & (0.179) \\ 
  & & & & \\ 
 Native speaker & 0.814$^{***}$ & 1.234$^{***}$ & 0.897$^{***}$ & 0.927$^{***}$ \\ 
  & (0.231) & (0.236) & (0.231) & (0.235) \\ 
  & & & & \\ 
 Dectect AI & $-$0.398 & $-$0.356 & $-$0.782$^{***}$ & $-$0.215 \\ 
  & (0.255) & (0.256) & (0.259) & (0.257) \\ 
  & & & & \\ 
 Jug favorable human & $-$0.917$^{**}$ & $-$1.484$^{***}$ & $-$0.507 & $-$1.091$^{***}$ \\ 
  & (0.418) & (0.430) & (0.433) & (0.420) \\ 
  & & & & \\ 
 AI knowledge & $-$0.150$^{*}$ & $-$0.070 & $-$0.130 & $-$0.045 \\ 
  & (0.084) & (0.081) & (0.082) & (0.081) \\ 
  & & & & \\ 
 AI usage daily & 0.116 & 0.180$^{**}$ & 0.158$^{**}$ & 0.144$^{**}$ \\ 
  & (0.075) & (0.073) & (0.072) & (0.073) \\ 
  & & & & \\ 
 AI usage work & $-$0.105 & $-$0.045 & $-$0.077 & $-$0.186$^{**}$ \\ 
  & (0.084) & (0.081) & (0.081) & (0.081) \\ 
  & & & & \\ 
 Amelioration & 0.174$^{*}$ & 0.129 & 0.225$^{**}$ & 0.149$^{*}$ \\ 
  & (0.091) & (0.090) & (0.088) & (0.089) \\ 
  & & & & \\ 
 Pertinence & $-$0.064 & $-$0.037 & $-$0.007 & $-$0.084 \\ 
  & (0.074) & (0.073) & (0.073) & (0.072) \\ 
  & & & & \\ 
 AI Creativity & $-$0.568$^{***}$ & $-$0.199 & $-$0.155 & 0.027 \\ 
  & (0.152) & (0.150) & (0.149) & (0.150) \\ 
  & & & & \\ 
 Willingness & 0.040 & 0.109 & $-$0.033 & $-$0.286$^{*}$ \\ 
  & (0.169) & (0.170) & (0.169) & (0.169) \\ 
  & & & & \\ 
\hline \\[-1.8ex] 
AIC & 3152.81 & 3219.35 & 3317.55 & 3330.93 \\ 
Observations & 975 & 975 & 975 & 975 \\
\hline 
\hline \\[-1.8ex] 
\end{tabular} 
\begin{tablenotes}
 \footnotesize
 \item \justifying   {\it Notes:}This table presents the coefficients reflecting the impact of evaluators' socio-demographic characteristics on the various creativity criteria scores. ***, ** and * indicate significance at the 1\%, 5\% and 10\% level, respectively. Effects are estimated using an Ordered Polynomial Logit. \end{tablenotes}
 \end{threeparttable}
 }
\end{table} 
\clearpage

\subsubsection{Comparison Between Agents per Criterion and Scores}\label{app:comp_top}

\begin{figure}[h!]
\centering
\caption{Comparison between Agents of Objective Measures}
\includegraphics[scale=0.078]{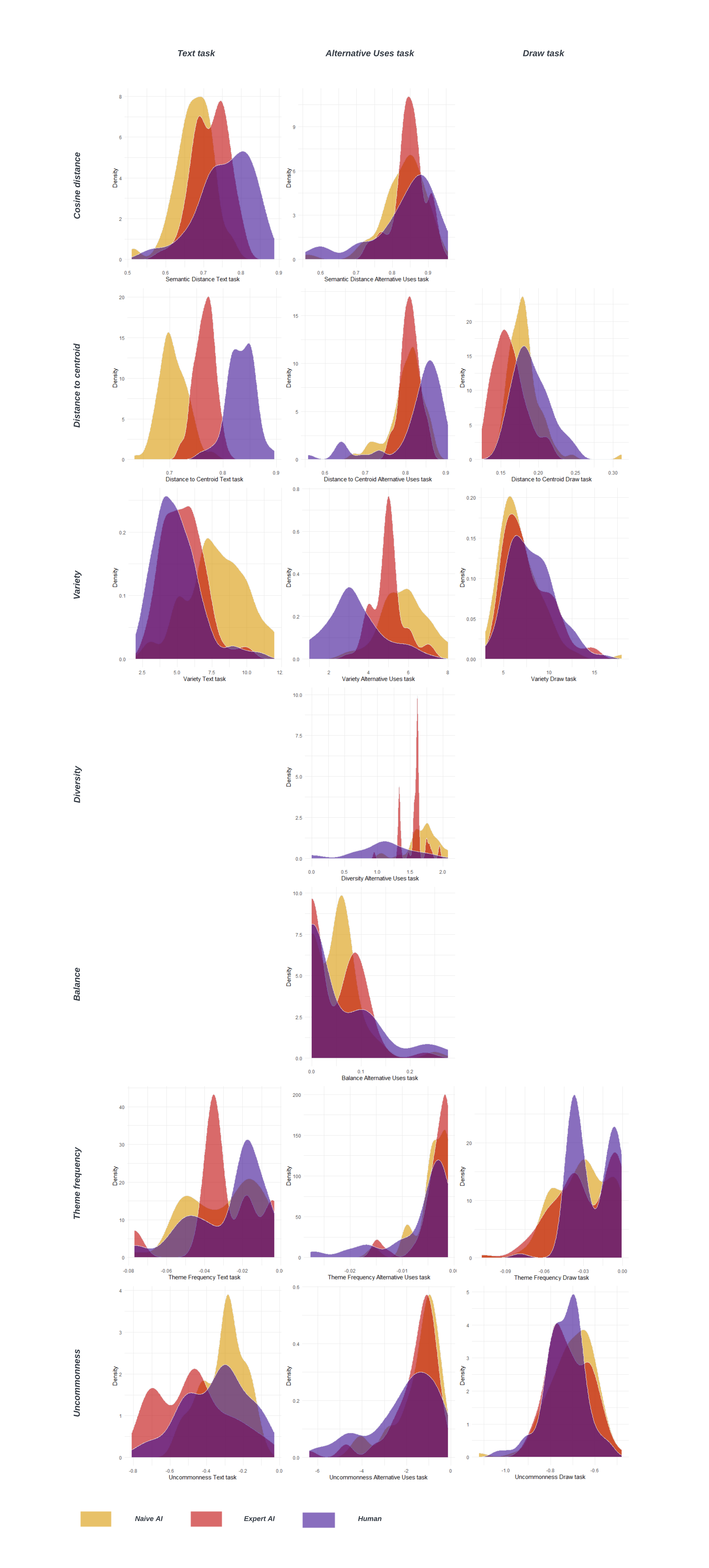}
\end{figure}

\begin{figure}[h!]
\centering
\caption{Comparison between Agents of Subjective Measures}
\includegraphics[scale=0.15]{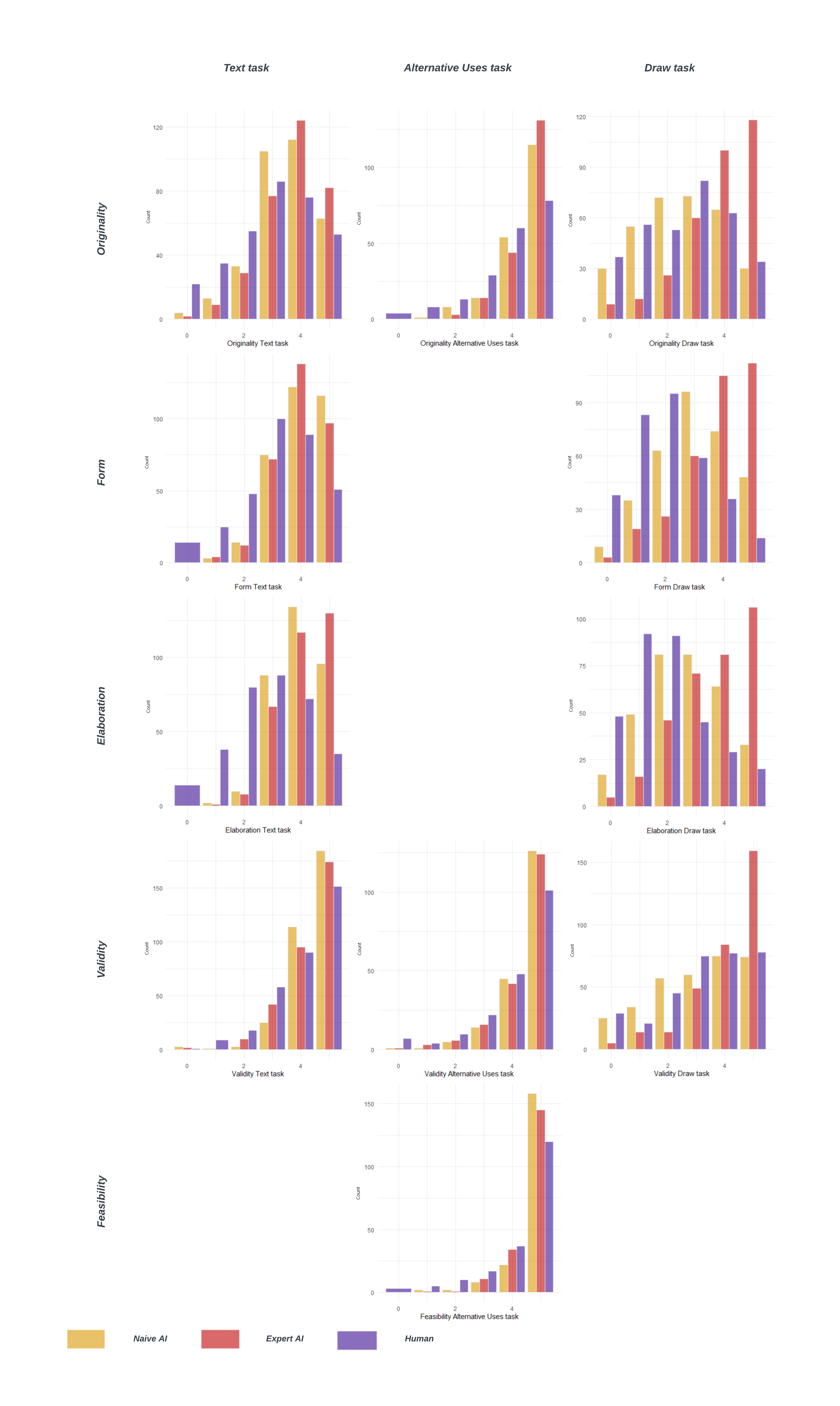}
\end{figure}

\end{document}